\documentclass[12pt]{article}

\usepackage{amssymb,amsmath,amsfonts}
\usepackage{authblk}
\usepackage{booktabs}
\usepackage{geometry,graphicx,color,setspace}
\usepackage[hidelinks]{hyperref}
\usepackage{etoolbox}
\usepackage{orcidlink}
\usepackage{rotating}

\def\sym#1{\ifmmode^{#1}\else\(^{#1}\)\fi}
\providecommand{\keywords}[1]
{
  \small	
  \textbf{\textit{Keywords---}} #1
}

\title{The AI research space shapes country academic performance}

\author[1]{Siyi Zhou\orcidlink{0000-0001-9540-9914}}
\author[2]{Tianxing Pan\orcidlink{0000-0002-6698-8387}}
\author[1]{Qing Ke\orcidlink{0000-0002-2945-5274}\thanks{q.ke@cityu.edu.hk}}

\affil[1]{Department of Data Science, College of Computing, City University of Hong~Kong, Hong Kong, China}
\affil[2]{School of Information Management, Nanjing University, Nanjing, China}

\begin{document}

\maketitle

\begin{abstract}
Recent advances in Artificial Intelligence (AI) have demonstrated great potential across diverse domains, prompting increasing attention to understanding the AI ecosystem. As knowledge production is often perceived as a process of connecting existing pieces of knowledge, mapping the interconnections of AI knowledge is crucial for unraveling the dynamics of scientific production in AI. However, maps created so far have fallen short of capturing knowledge relatedness, focusing instead on citation relationships between subfields. Here, we build an AI research space by connecting pairs of AI research topics based on their proximity measured through author-topic coupling. Projections of country level performances onto the space reveal that countries exhibit distinct baskets of specializations. We further examine the role of the structure of the AI research space in the scientific development of nations. We find that countries are more likely to gain specialization in topics that have high proximity to already specialized neighbor topics, and that publications whose topics align more with the developed topics in the country are more likely to become high impact. Our results highlight the AI development gaps across nations and the path-dependence in the development induced by the AI research space. 
\end{abstract}


\setcounter{page}{0}
\thispagestyle{empty}

\pagebreak \newpage

\doublespacing

\section{Introduction}

Artificial intelligence (AI), loosely referred to as technologies with complex, human-like reasoning capabilities, has a long history, dating back to the 1940s-50s when a number of ideas were introduced by scientists from several fields, including Wiener's cybernetics \cite{wiener1948cybernetics} and Pitts and McCulloch's networks of artificial neurons \cite{mcculloch1943logical}. Over the course of its development, there have been several streams of technologies used as manifestations of AI \cite{haenlein2019brief}. Most recent AI technology, as represented by machine learning in general and deep learning in particular, is a comeback of artificial neural networks and has become feasible only recently thanks to the availability of massive amounts of digital data fueled by powerful computing devices like GPU. Despite its recency, modern AI has been quickly realized for its tremendous implications for a series of domains, from medicine \cite{fleming2018artificial}, health care \cite{yu2018artificial}, and education \cite{popenici2017exploring} to transportation \cite{bojarski2016end}, finance \cite{dowling2023chatgpt}, and digital governance \cite{mondejar2021digitalization}. 

Prompted by the ubiquitous embrace, AI has been heralded as an example of general-purpose technologies (GPTs) \cite{bresnahan1995general, goldfarb2023could, eloundou2023gpts}, resembling the historical precedents such as steam engine and internal combustion engine. These are ``enabling technologies'', ``invention for the method of invention'', bearing the potential for pervasive applications in a wide range of sectors and enabling the generation of follow-up innovations that improve productivity in these applied sectors. The widespread interest in AI has also prompted studies that leverage scientific publications and patents to understand AI, including examining its GPT potential \cite{klinger2021deep, bianchini2022artificial}, labor force and productivity \cite{cockburn2018impact, van2020ai, yang2022artificial}, citation relationships \cite{frank2019evolution, klinger2020narrowing, gargiulo2023meso, alshebli2022beijing}, the role of industry \cite{ahmed2020democratization, martinez2021research, sevilla2022compute, ahmed2023growing}, talent mobility \cite{tang2020pace, jurowetzki2021privatization}, geographical factors \cite{bianchini2022artificial, alshebli2022beijing}, among many others \cite{krenn2023forecasting}. 

This study contributes to our understanding of the organization of the AI research ecosystem at a meso-scale. Existing studies to map its internal structure have typically relied on science classification schemes composed of subfields like computer vision (CV), natural language processing (NLP), robotics, etc. and examined knowledge flows between those subfields. However, AI research is a complex ecosystem, with its underlying knowledge components---techniques, tasks, and skills---displaying interconnections, and such complex intricacies spanning AI subfields have been largely ignored by current studies. For instance, although AI research fields may differ in applications, they may share similar foundational models or be highly correlated in addressing specific tasks \cite{raghu2020survey}. As an example, Transformers, a breakthrough neural network architecture, has been extensively incorporated in deep learning models in both CV and NLP. Providing a knowledge space of AI research would give us a more nuanced perspective about this domain that goes beyond the existing ones. 

Here we build an AI research space that captures proximity between topics. In doing so, we detect topics based on titles and abstracts of papers and find that there is a recent increase in the inequality in topic size, which coincides with the rise of deep learning. To construct the research space, we quantify the relatedness between two topics based on the similarity between their authors. Our inquiry into the structure of the research space indicates that there is a densely-connected core of topics that is also highly related to periphery topics. Furthermore, we project country level productivity onto the research space and find that countries possess differing specialization portfolios. For example, the US specializes in a diverse set of topics, China is largely over-represented in CV related topics, whereas Japan exhibits a strong focus on robotics. These patterns raise the question of how the structure of the AI research space shapes the scientific development of nations. We answer this question from two dimensions: topic transition and the production of high impact science. We find that, macroscopically, countries are more likely to gain specialization in topics that have high proximity to already specialized neighbor topics, and that, microscopically, publications whose topics align more with the developed topics in the country are more likely to become high impact. Our findings suggest that while knowledge diffusion is prevalent in the AI ecosystem, the path-dependent development induced by the structure of the AI research space still exists across nations.

\section{Data and method} \label{sec:method}

\subsection{Corpus of AI papers}

To compile our corpus of AI research papers, we begin by identifying relevant conferences and journals. For the former, we select the  conferences listed under the AI category recommended by the China Computer Federation (CCF)\footnote{\url{https://www.ccf.org.cn/Academic_Evaluation/By_category/}} and add the International Conference on Learning Representations (ICLR) to our list, resulting in 41 conferences. For journals, we choose the 146 ones listed under the ``Computer Science, Artificial Intelligence'' category in the Journal Citation Reports (JCR). We then search these venues in the Microsoft Academic Graph (MAG) \cite{sinha2015overview}, a database indexing over 220 million documents in a heterogeneous structure, and match 41 conferences and 139 journals, which in total publish $603,262$ papers. To ensure consistency across papers, we filter out workshop papers, as they typically enjoy a higher acceptance rate. We also exclude papers without abstract or author information, as well as papers published before 1990. These steps leave us with $496,577$ papers published between 1990 and 2021. Each paper in our dataset include valid metadata such as title, abstract, authors and their affiliations, document type, publication year, and citations. 

We present some summary statistics to better understand the characteristics of the corpus of AI papers. Overall, 50\% of the papers are conference proceedings. While the number of AI papers has surged over the past three decades, journal papers increase 18-fold, in contrast to 8-fold for conference papers (SI Fig.~S1A). Collaboration is on the rise: Single-author papers reduced from 31\% in 1990 to 3\% in 2021, accompanying the doubling of average team size from 2.1 to 4.2 authors per paper (SI Fig.~S1B). International collaboration has also increased steady, accounting for nearly 50\% of the papers in 2020 (SI Fig.~S3A). For US papers, there is an upward trend in the tendency to collaborate with China, whereas China's collaboration rate with the US has slightly declined since 2017 (SI Fig.~S3B). 

\subsection{Structural topic modelling of AI papers}

To construct our AI research space, we adopt topics as the base unit of knowledge component, given the intricate nature of the AI ecosystem, which comprises algorithms, terminology, tasks, etc. Topic modeling has been extensively used in innovation studies \cite{kaplan2015double}. However, traditional topic modelling techniques like the Latent Dirichlet Allocation (LDA) \cite{blei2003latent} make the exchangeability assumption, meaning that, for example, documents authored by one writer are exchangeable to another set of documents by a different author. Yet, existing practices tend to analyze topic difference by authors after fitting topic models, which violates the exchangeability assumption in the first place. To relax this assumption, a new method called structural topic modelling (STM) was proposed to take into account document level covariates \cite{roberts2014structural}. STM builds on top of the LDA and adds another component that connects document covariates to document topic distribution. Here we employ this technique to detect topics in our corpus of papers. 

Specifically, we derive a document-term matrix from the titles and abstracts, which are informative for paper content, and use this matrix as the input for STM. To extract a list of terms (\emph{i.e.}, vocabulary), we perform tokenization and lemmatization and extract meaningful phrases using AutoPhrase, a popular phrase extraction algorithm \cite{shang2018automated}. Stop words are not removed to avoid dropping potentially useful terms. In this step, $2,266$ papers without meaningful terms are removed, leaving us with $494,311$ papers as the final corpus. When fitting STM, we include years and document types (journal articles or conference proceedings) as the covariates, considering the substantial diversity in topics along the two dimensions. We set the number of topics, $K$, to $\{50,100,150,200,300,400,500\}$. We present our findings based on the model with $K = 150$, by balancing semantic coherence and exclusivity of the detected topics, a common method used for model selection in topic modelling (SI Fig.~S4). Our results remain robust when $K$ is 100 and 200 (see SI). For each topic, we manually assign a label based on its representative words.

\section{Results} \label{sec:result}

\subsection{Topics in AI research}

Before presenting our AI research space, we first present the results from STM, as they shed light on the changing landscape of AI research. We start by validating our topic modeling results. Firstly, we examine whether there are topic differences between journal and conference papers, which helps justify our choice of using document type as a covariate in STM. We perform, for each of the 150 topics, a regression analysis relating the proportion of the topic in a paper to its document type, finding that while journal articles emphasize topics like fuzzy logic and expert systems, conference papers are more likely to focus on robotics and various deep learning techniques (SI Fig.~S5). Secondly, we compare similarities between papers based on their topic distributions, expecting topic distinctions by AI subfields (\emph{e.g.}, CV and NLP). We find that this is indeed the case: For example, \emph{CVPR} papers are more similar to \emph{ICCV} papers than to \emph{EMNLP} or \emph{IROS} papers (SI Fig.~S6A); likewise, \emph{ACL} papers have higher similarities to \emph{EMNLP} papers than to \emph{ICCV} or \emph{IROS} papers (SI Fig.~S6B). However, similarity differences remain small, suggesting that the detected topics do not yield the same information as what is captured by publication venues. Indeed, the correlation between primary topics and venues is negligible (Spearman's rank correlation coefficient 0.05). 

\begin{figure*}[t!]
\centering
\includegraphics[trim=0 0 0 0, width=\textwidth]{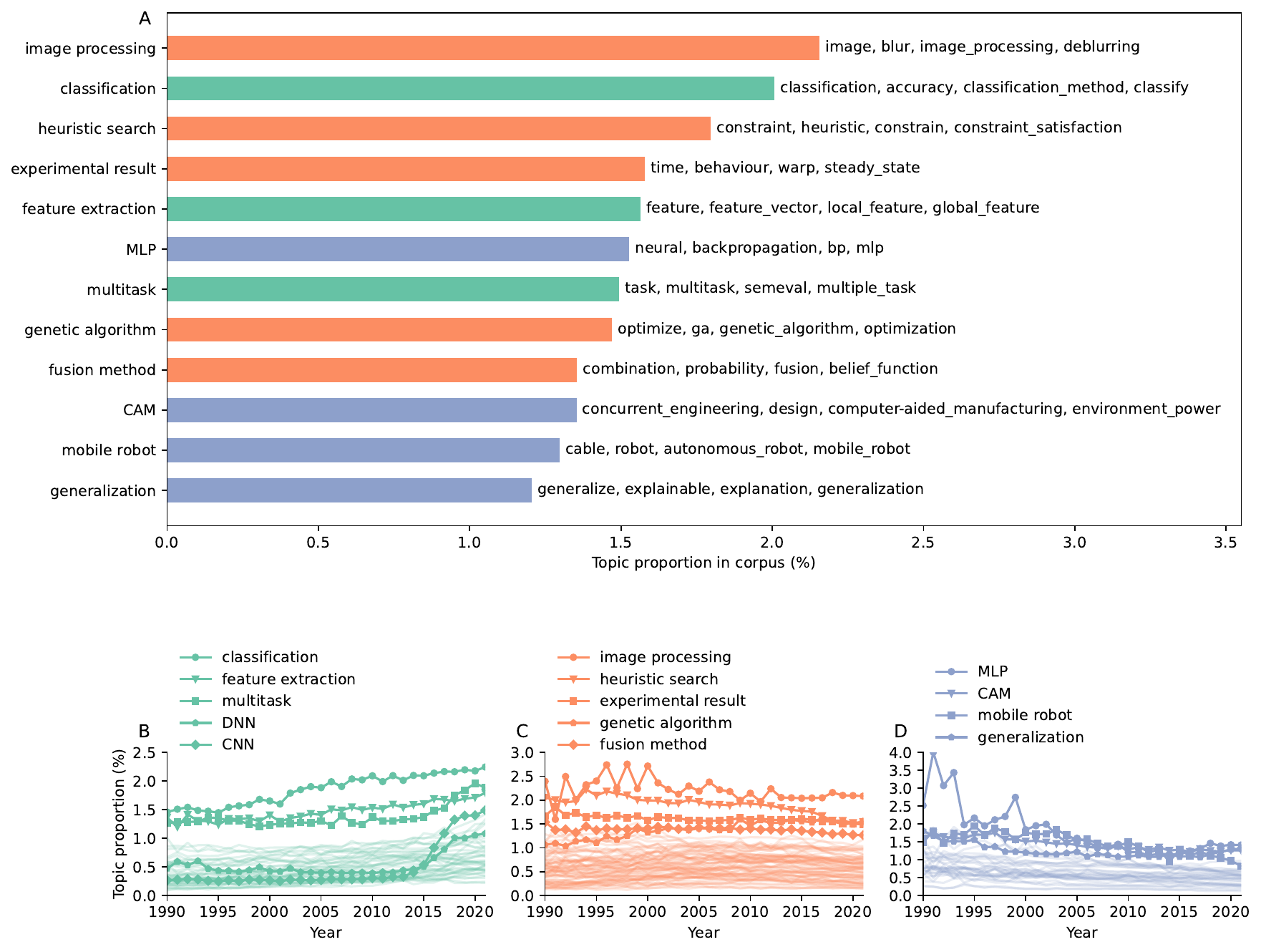}
\caption{Topics in AI research. 
(A) Proportions of the top 12 largest topics and their most representative phrases, selected based on their FREX scores, the product of frequency and exclusivity. SI Fig.~S7 shows the top phrases of other topics.
(B--D) Yearly normalized proportions of the 150 topics clustered into the (B) increasing, (C) stable, and (D) decreasing popularity categories. There are 45, 71, 34 topics in the 3 clusters, respectively. }
\label{fig:topic-year}
\end{figure*}

AI research encompasses topics with unequal presentations (Fig.~\ref{fig:topic-year}A). The most presented topic is image processing, which features words like ``blur'' and ``image''. Classification---one pillar of machine learning tasks---is the second largest topic, where words like ``accuracy'' and ``classification'' are representative ones. Other prominent topics include feature extraction, multilayer perceptron (MLP), fuzzy logic, graph neural network (GNN), and convolutional neural network (CNN) (see also SI Fig.~S7). 

As expected, AI research evolves rapidly, with its underlying topics exhibiting rich popularity dynamics (Figs.~\ref{fig:topic-year}B--D). We use the time series of topic proportions to cluster the 150 topics into 3 categories, corresponding roughly to rising, stable, and declining popularity. The group of topics with increasing popularity consists of foundational machine learning tasks such as classification and feature extraction. It also contains deep learning techniques like deep neural network and CNN, which have experienced a rapid increase since around 2012 (Fig.~\ref{fig:topic-year}B). The second group corresponds to topics with relatively stable popularity, including image processing, search, genetic algorithm, etc. (Fig.~\ref{fig:topic-year}C). The third category of topics displays decreasing popularity (Fig.~\ref{fig:topic-year}D). Notably, MLP---a forerunner of modern deep neural network---was the most focused topic in the early 1990s but lost its leading role gradually. 

Another trend behind the distributions of topic proportions over time is that the inequality in topic proportions, as measured by the Gini coefficient, kept decreasing until around 2012 and then started to increase (SI Fig.~S8). Such a U-shaped trend might be attributed to the rise of deep learning with growing dominance that overshadows other research topics, consistent with a previous finding concerning the narrowing of AI research directions \cite{klinger2020narrowing}. 

\subsection{The AI research space}

As mentioned, our aim is to build an AI research space that captures the shared capacities needed to produce AI knowledge, which enables our examinations of how countries ``learn''. To this end, in aligning with numerous prior inquiries \cite{guevara2016research, alabdulkareem2018unpacking, chinazzi2019mapping, hidalgo2007product}, we reason that shared capacities between topics are signaled by authors, as they are knowledge possessors. Specifically, if two topics share a higher similarity in their authorships, it indicates a larger possibility of authors publishing papers with both topics and thus suggests an easier transfer of knowledge between the two topics. Therefore, we define the relatedness between topics $k$ and $k'$, $\phi_{k,k'}$, as the cosine similarity between their author vectors, \emph{i.e.}, $\phi_{k,k'} = \cos \left(\mathbf{v}_k, \mathbf{v}_{k'}\right)$, where $\mathbf{v}_k$ is a row vector in the topic-author matrix that quantifies the factional count of papers belonging to topic $k$ and written by an author. 

\begin{figure*}[ht!]
\centering
\includegraphics[trim=0 0 0 0, width=\textwidth]{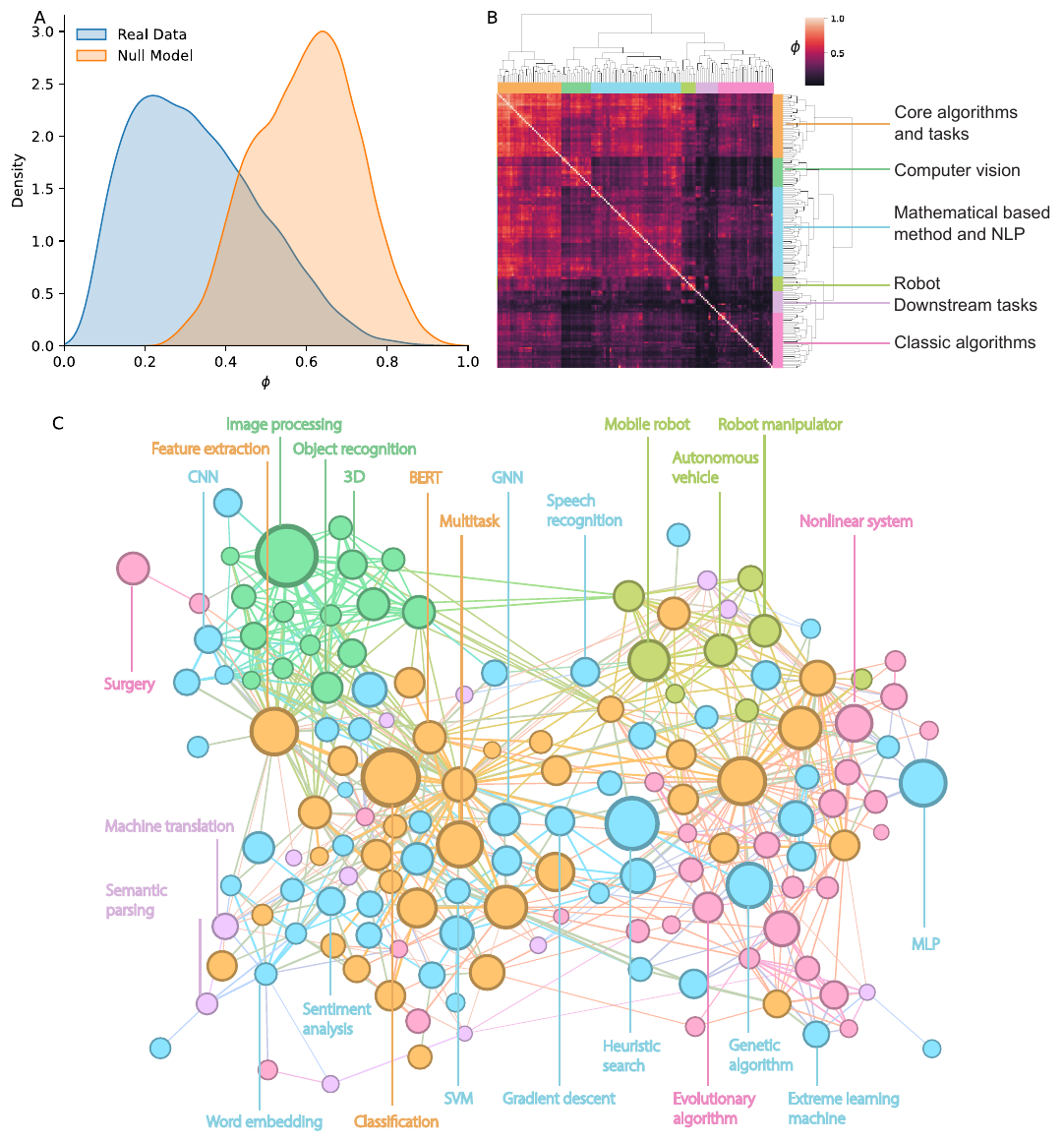}
\caption{The structure of the AI research space. 
(A) Distribution of observed relatedness $\phi$ between topics and the distribution from a null model, which is averaged over 1000 realizations. 
(B) Hierarchically clustered relatedness matrix between the 150 topics. 
(C) Backbone network representation of the AI research space \cite{serrano2009extracting}. Nodes represent AI topics, with size reflecting topic proportion in the corpus and color indicating cluster membership shown in B. Link thickness is proportional to $\phi$. SI Fig.~S11 presents the same network where all the nodes are labeled.}
\label{fig:topic-net}
\end{figure*}

The distribution of pairwise topic relatedness, shown in Fig.~\ref{fig:topic-net}A, indicates that topics share a limited relatedness (median $\phi_{k,k'}$ 0.31). It is also right-skewed, implying that a significant number of topic pairs have high proximity. To ascertain that our knowledge space does capture topic relatedness through author-topic coupling, we compare the empirical distribution with what would be expected from a null model, where we reshuffle the author lists of all the papers. The obtained null distribution is markedly different from the observed one, shifting to the right and skewed towards the left (Fig.~\ref{fig:topic-net}A). This is conceivable, as authors tend to work on a specific set of topics, and the random assignment of authors to papers destroys author-topic coupling, resulting in an enlarged set of topics for authors and consequently higher similarities between topics. 

To further reveal the structure of the AI research space, we perform hierarchical clustering of the $\phi$ matrix and identify six topic clusters (Fig.~\ref{fig:topic-net}B). The first one, referred to as the ``core algorithms and tasks'' cluster, includes topics such as classification, feature extraction, and BERT. They demonstrate strong correlations with both internal and external topics, highlighting their pervasiveness in AI research and thus distributed evenly in the central region in the network representation of the research space (Fig.~\ref{fig:topic-net}C). The second cluster, found in the upper left in Fig.~\ref{fig:topic-net}C, focuses on CV related topics like image processing, object recognition, 3D, among others. The third cluster is a large one consisting of math-related models like GNN, SVM (support vector machine), genetic algorithm, as well as certain NLP models and tasks like word embedding and sentiment analysis. These topics are distributed throughout the network, both in the central and peripheral regions. The fourth cluster is a small one dedicated to robotics related topics, including mobile robot, robot manipulator, and autonomous vehicle. The fifth cluster comprises downstream tasks located in the peripheral region of the network, including machine translation and semantic parsing. Finally, the last cluster consists of classic algorithms, such as nonlinear systems, evolutionary algorithms, and fuzzy logic, which are situated in the lower right in Fig.~\ref{fig:topic-net}C. SI Table~S1 lists all the topics in each cluster. 

\subsection{Country specialization and topic transition}

\begin{figure*}[t!]
\centering
\includegraphics[trim=0 5 0 0, width=.8\textwidth]{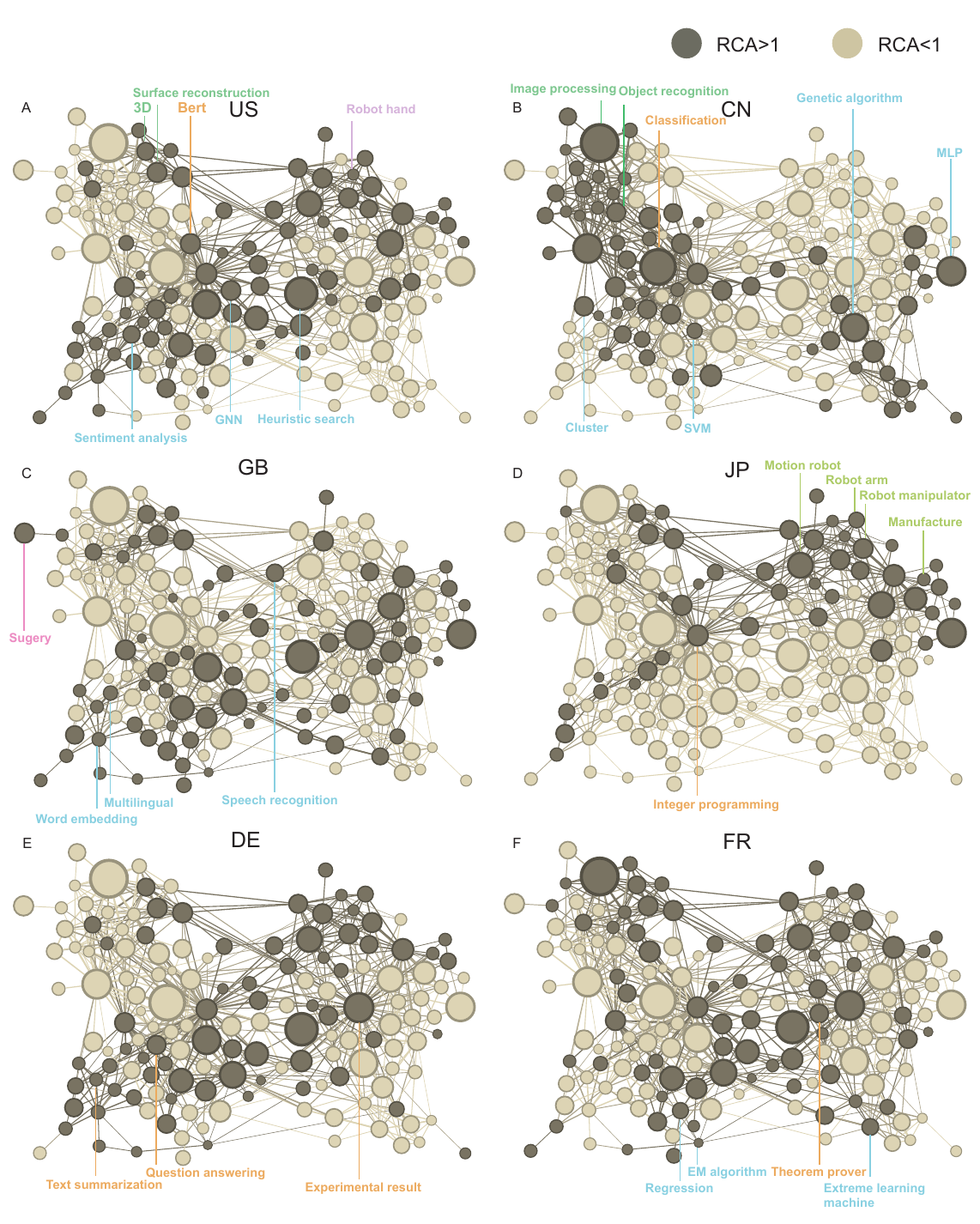}
\caption{Revealed comparative advantage in the 150 AI topics by country. (A) the US, (B) China, (C) the UK, (D) Japan, (E) Germany, and (F) France.}
\label{fig:topic-rca}
\end{figure*}

Having built the AI research space, we project country level performances onto the space. To quantify the performance of a country in a topic, we apply the revealed comparative advantage (RCA) measure, $r_{c,k}$, that has been widely used in a number of applications \cite{hidalgo2007product, hidalgo2009building, alabdulkareem2018unpacking}. RCA compares the productivity in topic $k$ with the global average, formally: 
\begin{equation}
r_{c,k} = \frac{x_{c,k}}{\sum_{k'} x_{c,k'}} \bigg/ \frac{\sum_{c'} x_{c',k}}{\sum_{c',k'} x_{c',k'}} \, ,
\end{equation}
where $x_{c,k}$ is the fractional number of papers produced from country $c$ with topic $k$. If $r_{c,k} > 1$, it means that country $c$ produces more papers with topic $k$, as a share of its total papers, than the global average, suggesting that this country exhibits specialization in the topic. 

We calculate RCA for the 6 most prolific countries in AI research, namely the US, China, the UK, Japan, Germany, and France. They collectively account for 60\% of the papers in our corpus. The results reveal distinct specialization patterns across countries (Fig.~\ref{fig:topic-rca}). To elaborate, the US specializes in a diverse set of topics, particularly those related to NLP and robotics, such as BERT, mobile robot, and multimodal. By comparison, China's specialized topics predominantly lie in the two regions in the research space: one in the upper left encompassing CV related topics like image processing and object recognition, and the other in the lower right emphasizing classical statistical methods such as clustering and SVM. The UK displays sparse specializations in NLP and classic algorithms, whereas Japan excels narrowly in robotics and manufacturing related topics. Germany, similar to the US, prominently shows specialization in NLP. Finally, France is over-represented primarily in mathematical topics. 

These aggregated patterns mask temporal variations in a country's RCA performance. For instance, SI Fig.~S12 presents how the baskets of specialized topics have changed during the three decades in 1990s--2010s for the US and China. It shows that China already gained specializations in CV topics in the 2000s, whereas the US presented its advantage in NLP topics only in the past decade. 

\begin{figure*}[t!]
\centering
\includegraphics[trim=0 6 0 0, width=\linewidth]{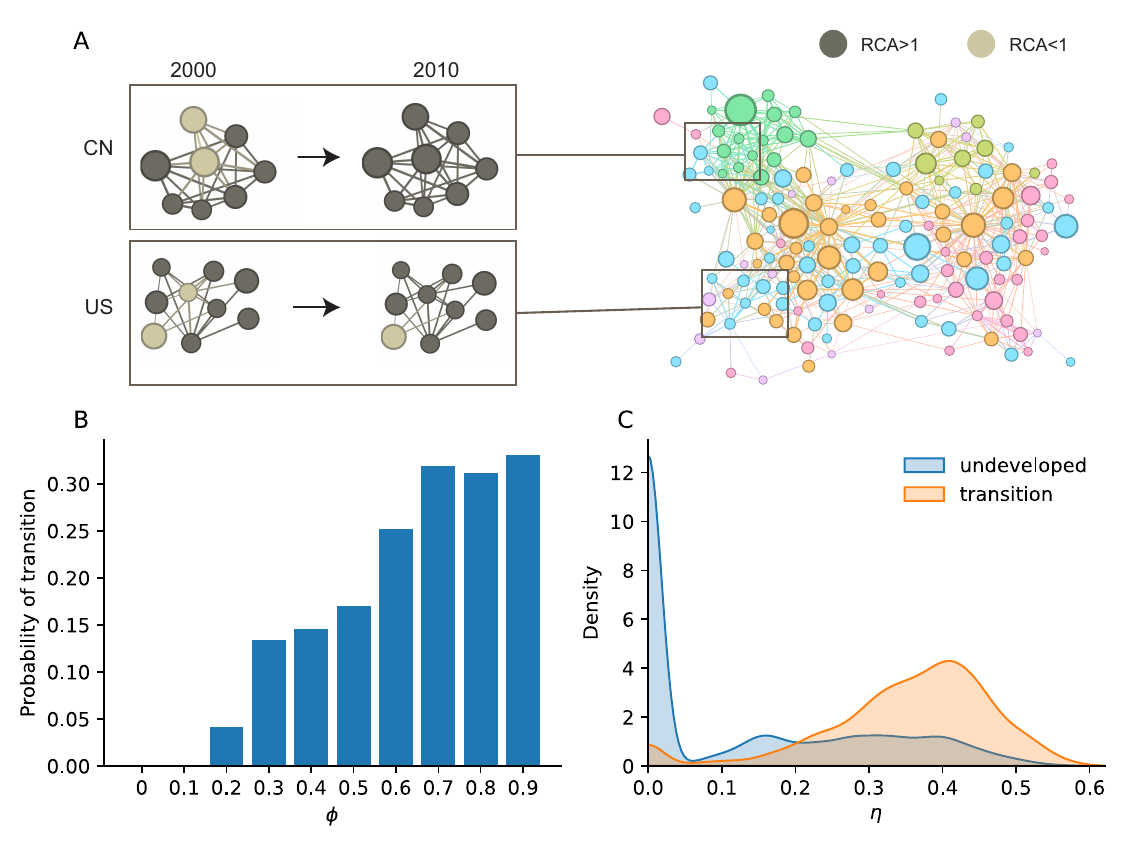}
\caption{The role of related topics in the development of specialization.
(A) Illustrations of two transition topics.
(B) The dependence of transition on the maximum relatedness to other developed topics. 
(C) Distributions of density scores, $\eta$, for transition and undeveloped topics.}
\label{fig:topic-transition}
\end{figure*}

In general, we seek to understand what factors contribute to a country's development of expertise in a specific AI topic. Inspired by an burgeoning stream of literature on the principle of relatedness stating that empirically the probability of a region entering an economic activity is associated with the number of related activities located there \cite{hidalgo2018principle}, here we similarly investigate how specialized neighbor topics in the AI research space constrain the development of specialization. To this end, we divide the focused period into three separate decades: 1990s (denoted as $d=0$), 2000s ($d=1$), and 2010--2021 ($d=2$), and define $e_{c,k}^d = 1$ if $r_{c,k}^d > 1$, and $e_{c,k}^d = 0$ otherwise. Then, for a topic $k$ with $e_{c,k}^d = 0$, we classify it as a transition topic if it satisfies $e_{c,k}^{d+1} = 1$, and an undeveloped topic otherwise. 

We present two pieces of evidence supporting that topics experiencing transition are surrounded by numerous developed topics, as illustrated in Fig.~\ref{fig:topic-transition}A. First, for a topic $k$, the probability of transitioning increases with the maximum relatedness $\phi_{k,k'}$ among $k$'s nearby developed topics (Fig.~\ref{fig:topic-transition}B). Second, we quantify the extent of development of neighbor topics using a measure called density~\cite{hidalgo2007product}. It is defined as, for country $c$ in topic $k$, the fraction of total relatedness to other developed topics, formally: 
\begin{equation}
\eta_{c,k}^d = \sum_{k'} e_{c,k'}^{d-1}\phi_{k,k'} \bigg/ \sum_{k'}\phi_{k,k'} \, .
\end{equation} 
Fig.~\ref{fig:topic-transition}C shows distinct distributions of $\eta$ for transition topics and undeveloped ones, with transition topics exhibiting much larger density scores than undeveloped topics. This indicates that the more relatedness-weighted developed topics surrounding an undeveloped topic, the higher chance it will be developed in the next decade. These findings provide additional empirical evidence for the principle of relatedness and point to the existence of constraints imposed by the structure of knowledge space on the development of specialization. Namely, developing expertise in AI by a nation may not be instantaneous but follow a gradual and sequential trajectory towards existing developed topics. Such path-dependent development may also suggest that enhancing the development level of nearby topics could be a beneficial strategy for countries aiming to promote certain topics. 

\subsection{Production of high-impact research} \label{subsec:impact}

\begin{figure*}[t!]
\centering
\includegraphics[trim=4 4 4 0, width=\linewidth]{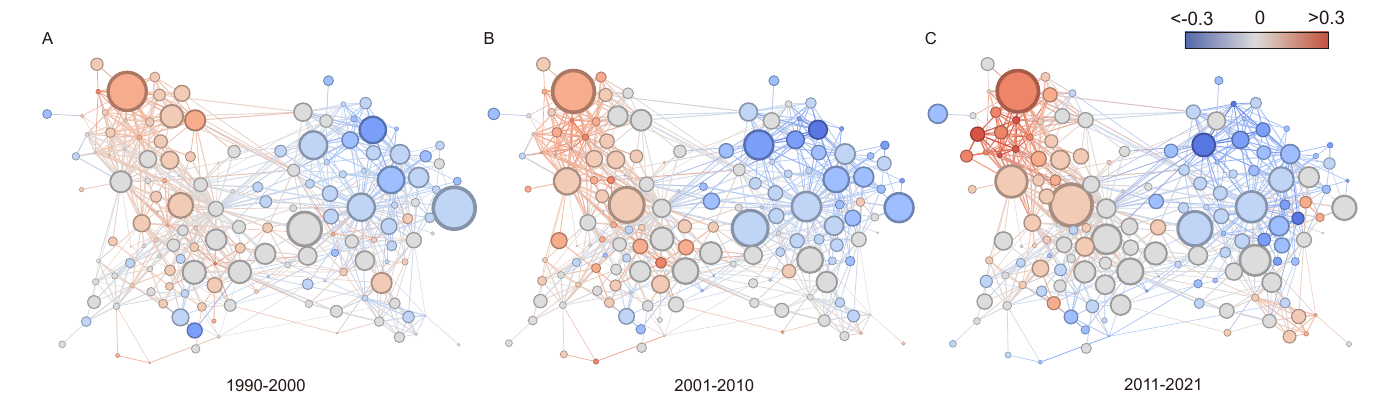}
\caption{Empirical conditional probability, by decade, of finding a high impact paper in one topic.}
\label{fig:topic-impact}
\end{figure*}

Our results so far suggest that the structure of the AI research space influences macroscopically the AI research development of nations. In this section, we ask whether it also impacts microscopically the production of high impact science from a country. We first project high impact papers (5-year citations ranked into the top 5\% among papers published in one decade) onto the research space, by calculating the probability of a paper being high impact, conditional on the fact that the paper proportionally belongs to the topic. The results, which are presented in Fig.~\ref{fig:topic-impact}, indicate that there are ``hotspots'' (``coldspots'') in the space where it is more (less) likely to identify high impact papers. For instance, topics related to CV, such as image processing, are more likely to feature highly cited papers and the tendency has been reinforced over time. Many robotics related topics, on the other hand, are less likely to produce high impact papers. These results indicate that the AI research space is an influential factor. 

To quantify the role of the research space for different countries, we propose a measure called adjacent development level, $\omega_{i,c}$, that evaluates to what extent a paper $i$'s topic distribution aligns with the developed topics in the country $c$ where it is produced. Formally, it is density weighted sum of its topic distribution: 
\begin{equation}
\omega_{i,c}^d = \sum_{k} \eta_{c,k}^{d-1} \cdot p_{i,k} \, ,
\end{equation}
where $p_{i,k}$ is the proportion of topic $k$ in paper $i$. For papers contributed by more than one country, we take the average of $\omega_{i,c}$ (see SI Fig.~S14 for the distributions of $\omega_{i,c}$ by country). Our hypothesis is that $\omega_{i,c}^d$ is positively correlated with the probability of $i$ being high impact. To test this, we define high impact AI papers if their five-year citations are ranked into the top 5\% among papers grouped by publication year. SI Fig.~S15A shows empirical probabilities of being high impact for different groups of papers based on their $\omega$ values, confirming a positive association; when adjacent development level $\omega_{i,c}^d$ is approximately 0.6, the probability of becoming high impact is four times higher compared to when $\omega_{i,c}$ is around 0.1. 

\begin{table}[t!]
\centering
\caption{Logistic regression modeling of high impact papers.}
\label{tab:logit}
\begin{tabular}{l*{2}{c}}
\hline\hline
                            &\multicolumn{1}{c}{(1)}&\multicolumn{1}{c}{(2)}\\
\hline
Authors                     &      0.0658\sym{***}&      0.0646\sym{***}\\
                            &   (0.00508)         &   (0.00506)         \\
International collaboration &       0.387\sym{***}&       0.214\sym{***}\\
                            &    (0.0210)         &    (0.0213)         \\
Company coauthorship        &       0.425\sym{***}&       0.295\sym{***}\\
                            &    (0.0254)         &    (0.0257)         \\
Adjacent development level  &                     &       5.074\sym{***}\\
                            &                     &     (0.128)         \\
Constant                    &      -2.401\sym{***}&      -4.551\sym{***}\\
                            &    (0.0857)         &     (0.102)         \\
\hline
Year fe                     & $\checkmark$        & $\checkmark$        \\
Field fe                    & $\checkmark$        & $\checkmark$        \\
Observations                &      279475         &      279475         \\
Pseudo \(R^{2}\)            &       0.031         &       0.046         \\
\textit{BIC}                &    112976.3         &    111256.3         \\
\hline\hline
\multicolumn{3}{l}{\footnotesize Standard errors in parentheses}\\
\multicolumn{3}{l}{\footnotesize \sym{*} \(p<0.05\), \sym{**} \(p<0.01\), \sym{***} \(p<0.001\)}\\
\end{tabular}
\end{table}

\begin{figure}[t!]
\centering
\includegraphics[trim=0 2 0 0, width=.5\textwidth]{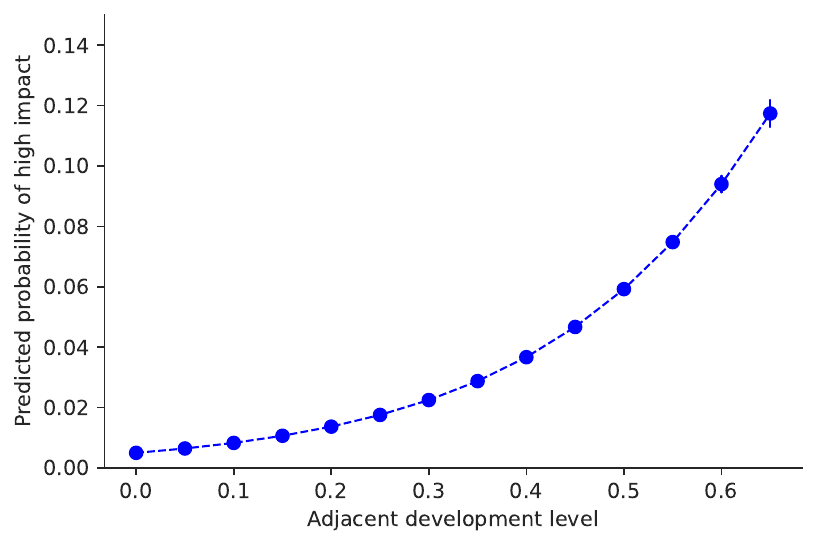}
\caption{Predicted probability of being high impact paper increases with adjacent development level, indicating that papers whose topic distribution aligns more with the developed topics in the country are more likely to become high impact.}
\label{fig:paper-impact}
\end{figure}

To further quantify the effect of adjacent development level on the production of high impact papers, we fit logistic regression models while controlling for other confounders. First, we control for number of authors. Second, we include an indicator encoding whether a paper involves international collaboration \cite{martinez2021research}. Third, we consider a binary variable for industry collaboration, as inferred from affiliations, given the increasingly indispensable role of large data and computing resources in producing high impact AI research which can be readily provided by companies \cite{ahmed2023growing}. Fourth, we include publication year as a fixed effect to control for longitudinal changes, such as global scientific investment, that may affect knowledge production. Finally, we include subfield designation as another fixed effect to account for its impact on citations. Subfields of papers are taken as the fields of study (FoS) located at the second level in the FoS hierarchy provided in the MAG, containing areas like pattern recognition, computer vision, etc. Table~\ref{tab:logit} presents the detailed regression results. It indicates that the likelihood of being a high impact paper is associated with numerous factors. Papers resulted from international collaboration have a $24\%$ increase in the odds of being high impact. The role of collaboration with industry is larger; papers with company-affiliated coauthors have a 34\% increase in the odds. After controlling for all the considered factors, a one standard deviation increase in adjacent development level translates to 1.8 times standard deviation increase in the log odds of being high impact, with the predicted probability plotted in Fig.~\ref{fig:paper-impact}. These results suggest that publications whose topics align more with the developed topics in the country are more like to become high impact.

\section{Discussion}

Using about half a million AI papers published in over three decades, we have constructed an AI research space that captures the relatedness between AI topics.  It allowed us to fingerprint specialization portfolios of countries and further examine how its structure shaped the national development of AI research and affected the production of high impact science. Our work thus contributes to two lines of literature. First, we contribute to the literature of relatedness and diversification \cite{hidalgo2018principle}. Previous inquiries have revealed research space within physics \cite{chinazzi2019mapping, tripodi2020knowledge}, global science \cite{guevara2016research, miao2022latent, alshamsi2018optimal, galuppo2021evaluating}, and technological areas \cite{breschi2003knowledge}. We join these studies by presenting a research space in AI. Moreover, our results pointed out the path-dependence of AI development, similar to the development of many other areas. Second, by exploring the structure of the research space, we join the burgeoning literature of understanding the AI ecosystem and its impact on scientific innovation. Finally, although our focus is on AI, our framework can be extended to other areas of science to uncover underlying topic clusters within their knowledge space and gain insights into the interconnectedness within those fields. 

Policy implications emerged from our findings worth discussions. The research space we built may help policymakers identify the positions of their countries in the map, which may have further implications for relevant stakeholders in making informed decisions about where to invest in AI research based on location-specific research strengths and target areas to excel. Having said so, it is worth to mention that although endogenous scientific capacity building is important, curiosity-driven exploration remains still as a leading way for knowledge emergence and science advances. Thus, a delicate balance between mission-oriented capacity building efforts and curiosity-driven research should be pursued when it comes to public funding allocation. 

Our work has the following limitations. First, it is limited by the focused corpus of AI papers. Specifically, we considered papers published until 2021, thus our analysis was not able to reflect the most recent development in this rapidly-changing field. Similarly, our venue-based criterion to include papers means that we missed AI research published in other venues. Of a related note, defining AI and AI research is a Herculean task, considering its long history and ubiquitous applications. Yet, it has a direct implication on the identification of study samples. Existing studies---including ours---have been relatively silent on the development of a consensual definition of AI research. Second, we derived topics based solely on title and abstract, ignoring other aspects like citation relationships between documents that are also informative for topic detection. Third, our constructed research space is static, while in reality it evolves over time with new research directions emerging. 

Despite these limitations, the availability of a research space enables many lines of inquiry for future work. For instance, one could study how AI researchers move in the intellectual space with what career outcome. Second, we have studied the role of the research space on varying rates of AI development across countries and their over-representation in different knowledge components. Future work may explore other factors such as research funding, industrial policy, talent mobility, and academia-industry collaboration. These lines of studies would gain a deeper understanding of the dynamics shaping the global landscape of research and development of AI---technologies that are widely perceived to have far-reaching implications. 

\section*{Acknowledgment}

This work was partially supported by the National Natural Science Foundation of China (72204206), City University of Hong Kong (Project No. 9610552, 7005968), and the Hong Kong Institute for Data Science. S.Z. was supported by The Laboratory for AI-Powered Financial Technologies Limited (AIFT).

\clearpage

\setcounter{figure}{0}
\renewcommand{\thefigure}{S\arabic{figure}}

\setcounter{table}{0}
\renewcommand{\thetable}{S\arabic{table}}

\setcounter{equation}{0}
\renewcommand{\theequation}{S\arabic{equation}}

\section*{Supporting Information} \label{sec:si}

\addcontentsline{toc}{section}{Supporting Information}

\begin{figure}[ht!]
\centering
\includegraphics[trim=0 5 0 0, width=\linewidth]{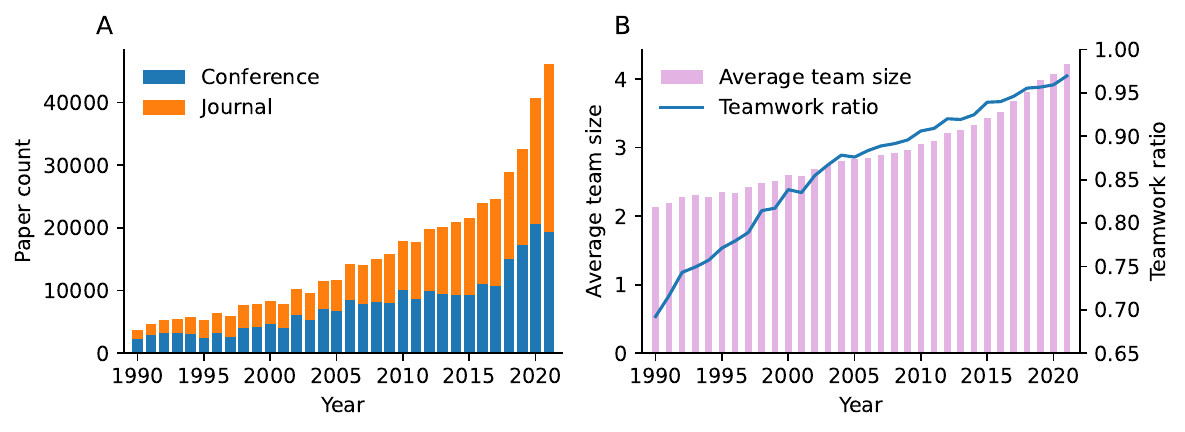}
\caption{Trends in AI paper count, teamwork ratio, and team size. 
(A) AI papers surged over the past three decades, with journal papers increasing 18-fold and conference papers 8-fold from 1990 to 2021. 
(B) Single-author AI papers reduced to 3\% in 2021 from 31\% in 1990, while the average team size doubled from 2.1 to 4.2 authors per paper.}
\label{fig:si:paper count}
\end{figure}

\begin{figure}[ht!]
\centering
\includegraphics[trim=0 7 0 0, width=\linewidth]{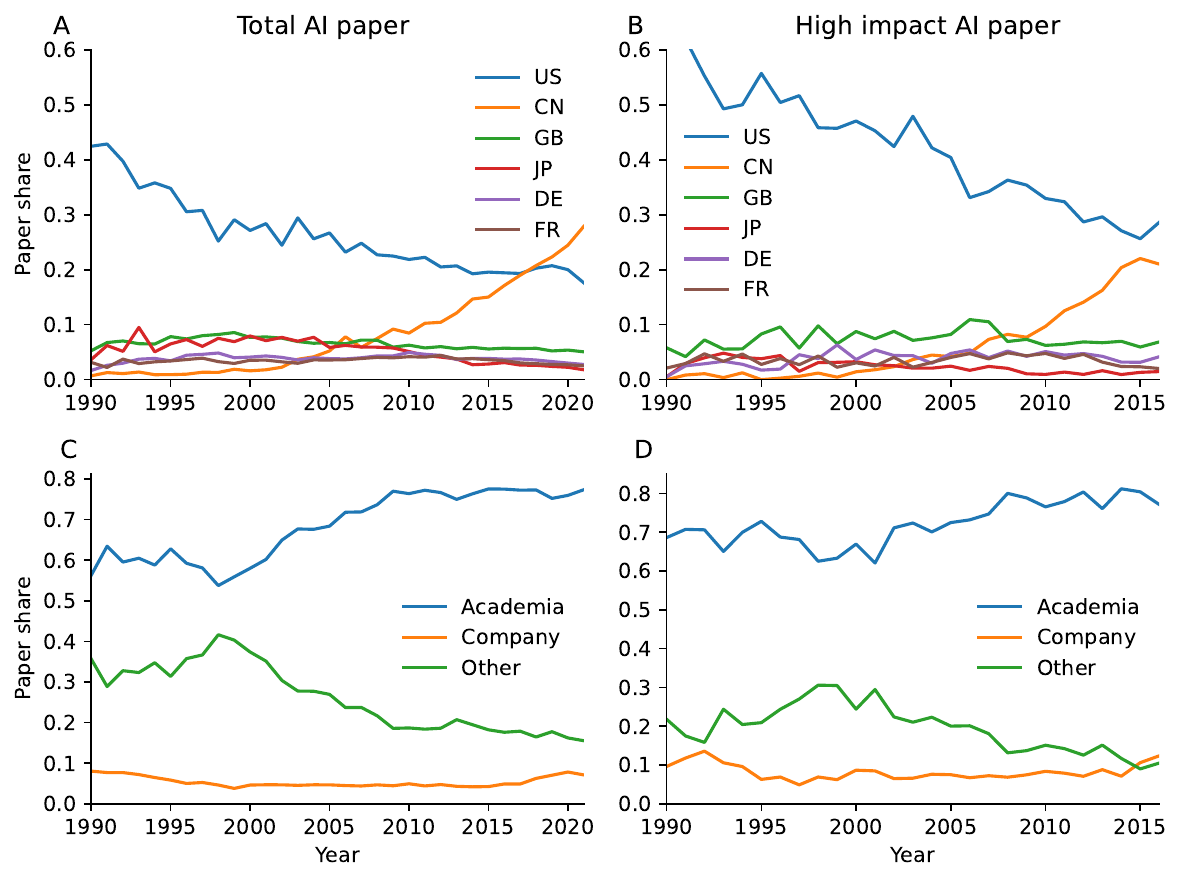}
\caption{Trends in authorship of AI papers. 
(A--B) The 6 most productive countries contribute to over 60\% of global AI papers. China surpassed the US in AI paper share in 2018, but the US still led in high impact papers in 2016 (28\%). 
(C--D) While academic institutions dominate AI paper production, company-affiliated papers are more likely to achieve high impact.}
\label{fig:si:paper share}
\end{figure}

\begin{figure}[t!]
\centering
\includegraphics[trim=0 12 0 0, width=\linewidth]{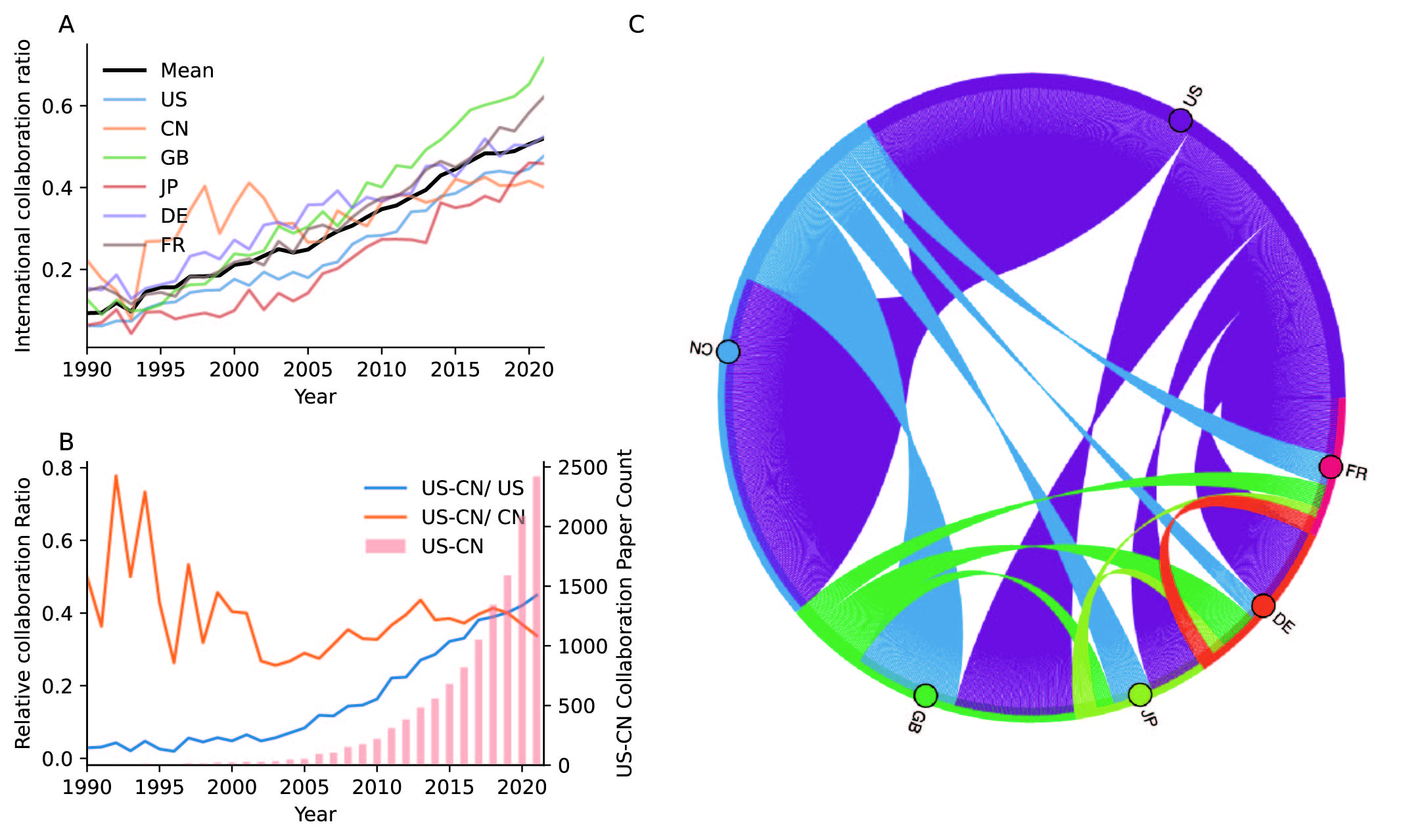}
\caption{International collaboration in AI papers. 
(A) International collaboration rates of the 6 focused countries, showing that Japan and the US have lower rates than the global mean. 
(B) The number of AI papers resulted from collaborations between the US and China steadily increase over time, with the US showing an upward trend in the tendency to collaborate with China. However, China's collaboration rate with the US has slightly declined since 2017. 
(C) Overall collaborations in AI papers between the 6 countries.}
\label{fig:si:intl}
\end{figure}

\begin{figure}[t!]
\centering
\includegraphics[trim=0 8 0 0, width=\linewidth]{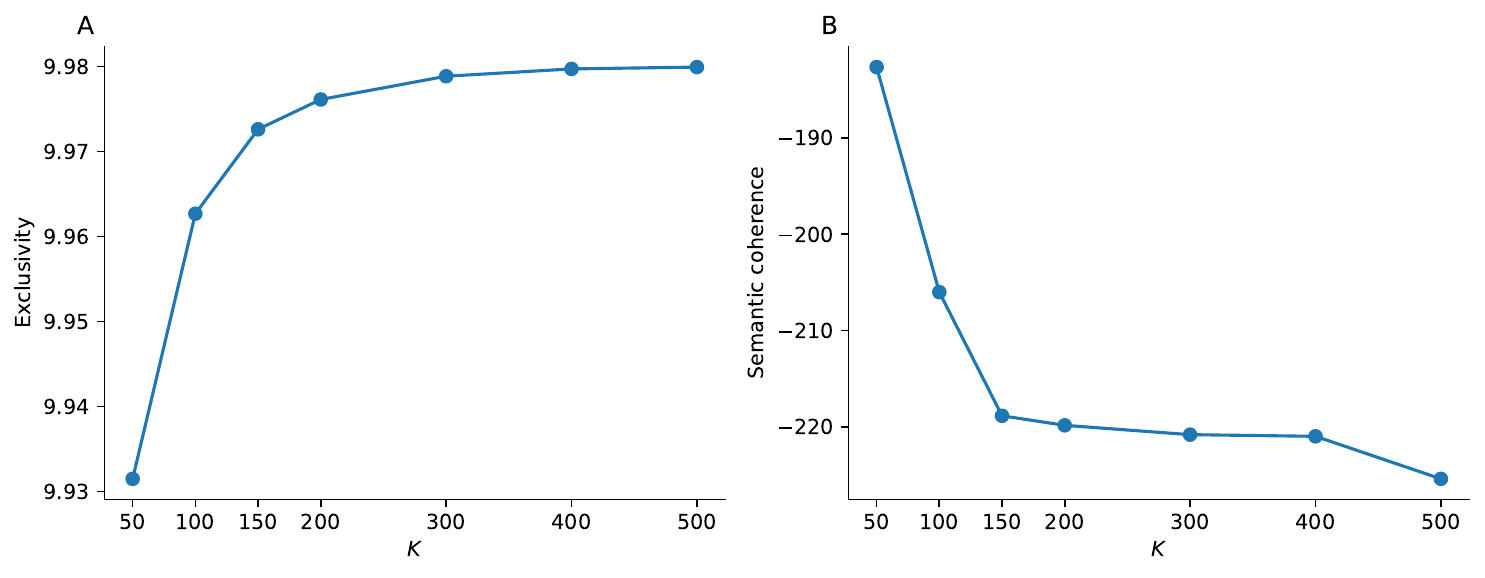}
\caption{Model selection of topic modeling.
(A--B) Topic exclusivity and semantic coherence for different $K$, number of topics. We select $K=100,150,200$ for our further analysis to balance exclusivity and coherence.}
\label{fig:si:model-selection}
\end{figure}

\begin{figure}[t!]
\centering
\includegraphics[trim=0 6 0 0, width=.95\linewidth]{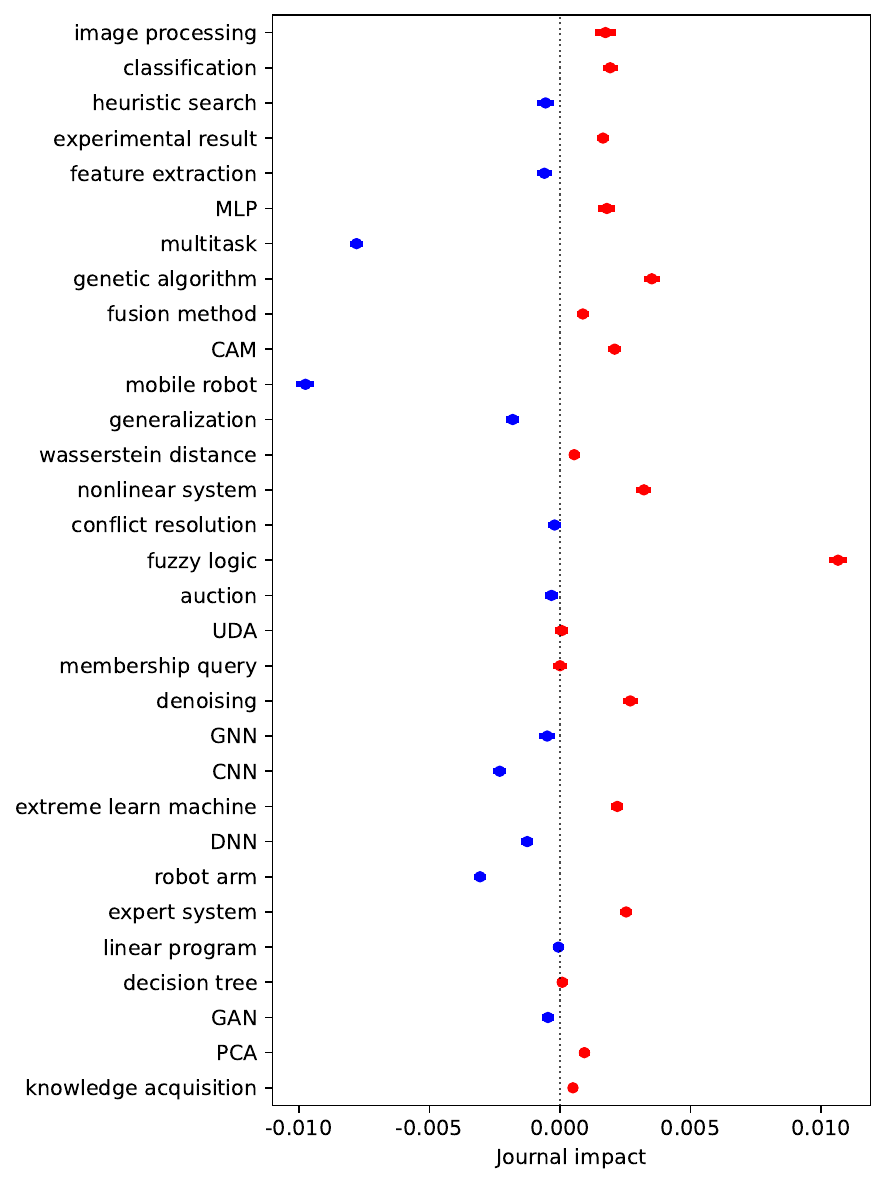}
\caption{The impact of document type (journal vs. conference paper) on topic prevalence.
For each topic, we perform a linear regression where the dependent variable is topic prevalence and the independent variable is document type. Red dots represent a positive journal impact on the focal topic prevalence, whereas blue dots represent negative impact.}
\label{fig:si:prevalence}
\end{figure}

\begin{figure}[t!]
\centering
\includegraphics[trim=0 4 0 0, width=\linewidth]{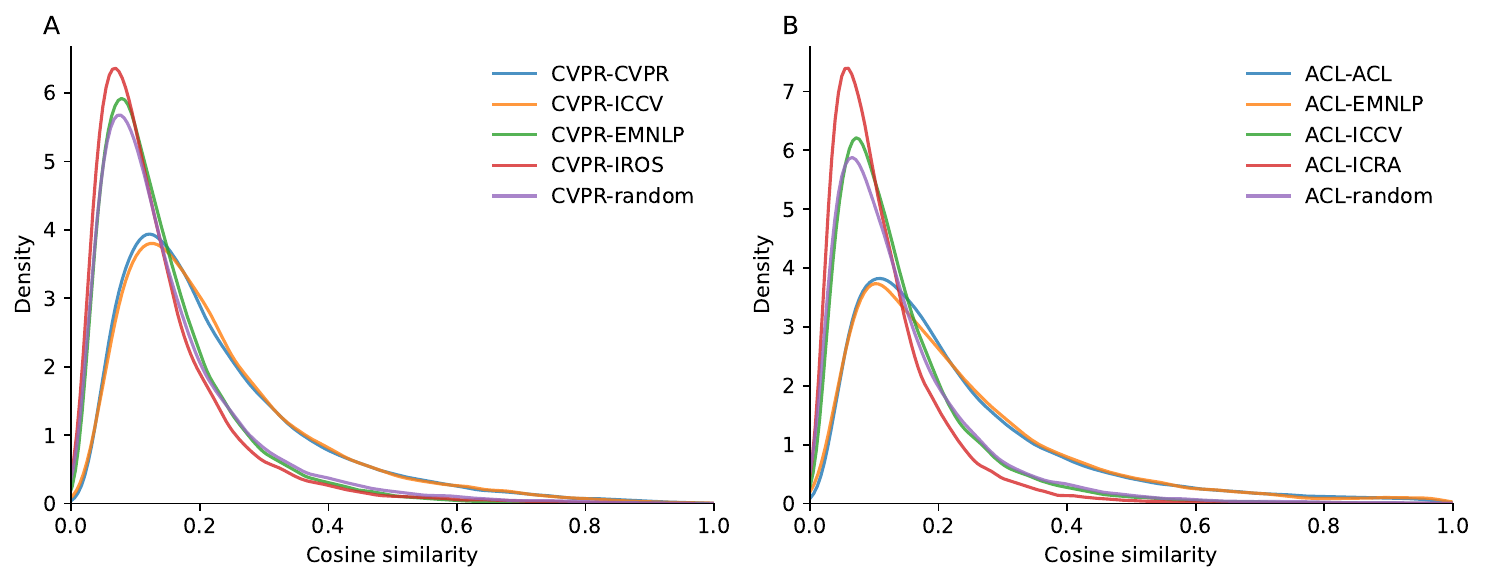}
\caption{Validation of topic modeling.
(A--B) Distributions of cosine similarities between papers from different sets. 
Using pairwise cosine similarity between two randomly sampled papers within the same year, papers from the same conference or conferences in the same subfield have a higher mean similarity (around $0.23$). In comparison, papers from conferences in different subfields show lower mean similarity (around $0.14$). This finding holds for both computer vision (A) and natural language processing (B).}
\label{fig:si:paper-sim}
\end{figure}

\begin{figure}[t!]
\centering
\includegraphics[trim=0 10 0 0, width=\linewidth]{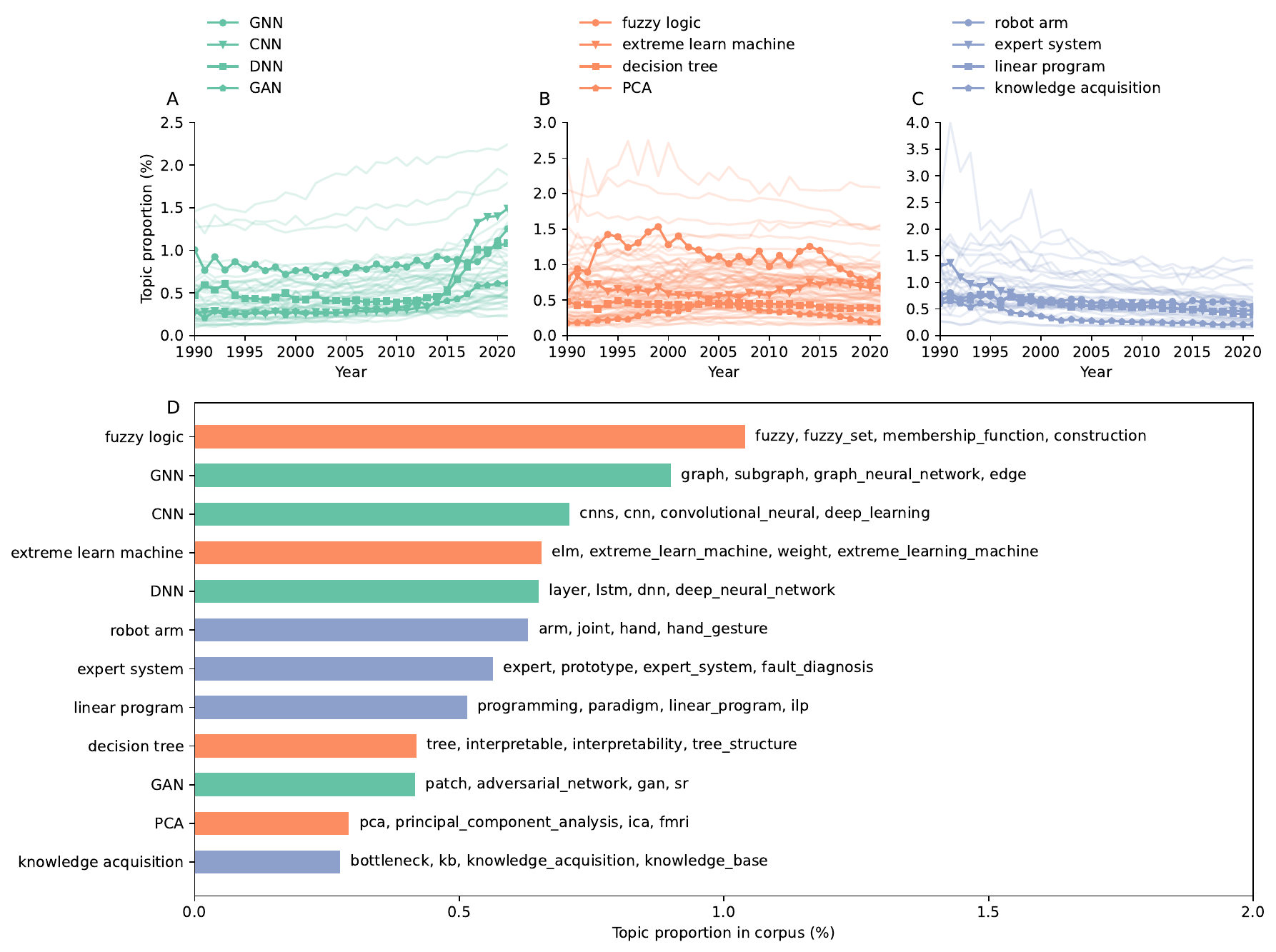}
\caption{Topics in AI research. 
The same as in Fig.~1 in the main text, but with different sets of topics highlighted.}
\label{fig:si:ai_topic}
\end{figure}

\begin{figure}[t!]
\centering
\includegraphics[trim=0 8 0 0, width=\linewidth]{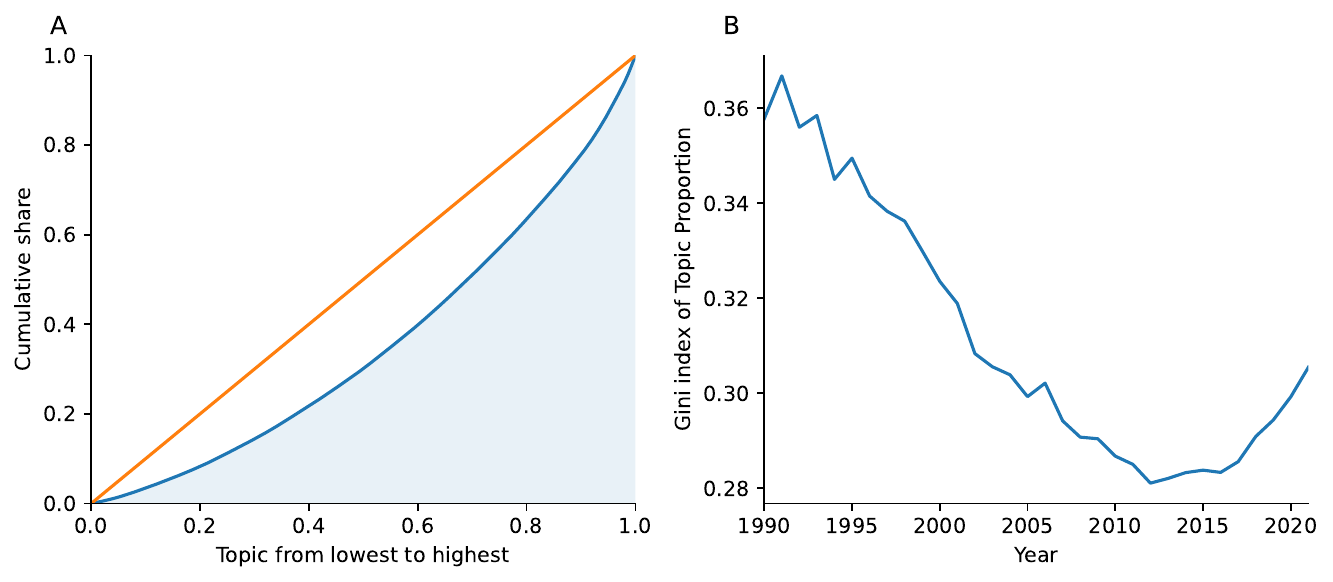}
\caption{Inequality in AI topics prevalence.
(A) The Lorenz curve for overall topic proportions.
(B) Gini coefficient of topic proportions over the past three decades, indicating a recent increase in the inequality in AI topics prevalence.}
\label{fig:si:gini}
\end{figure}

\begin{figure}[t!]
\centering
\includegraphics[trim=0 6 0 0, width=\linewidth]{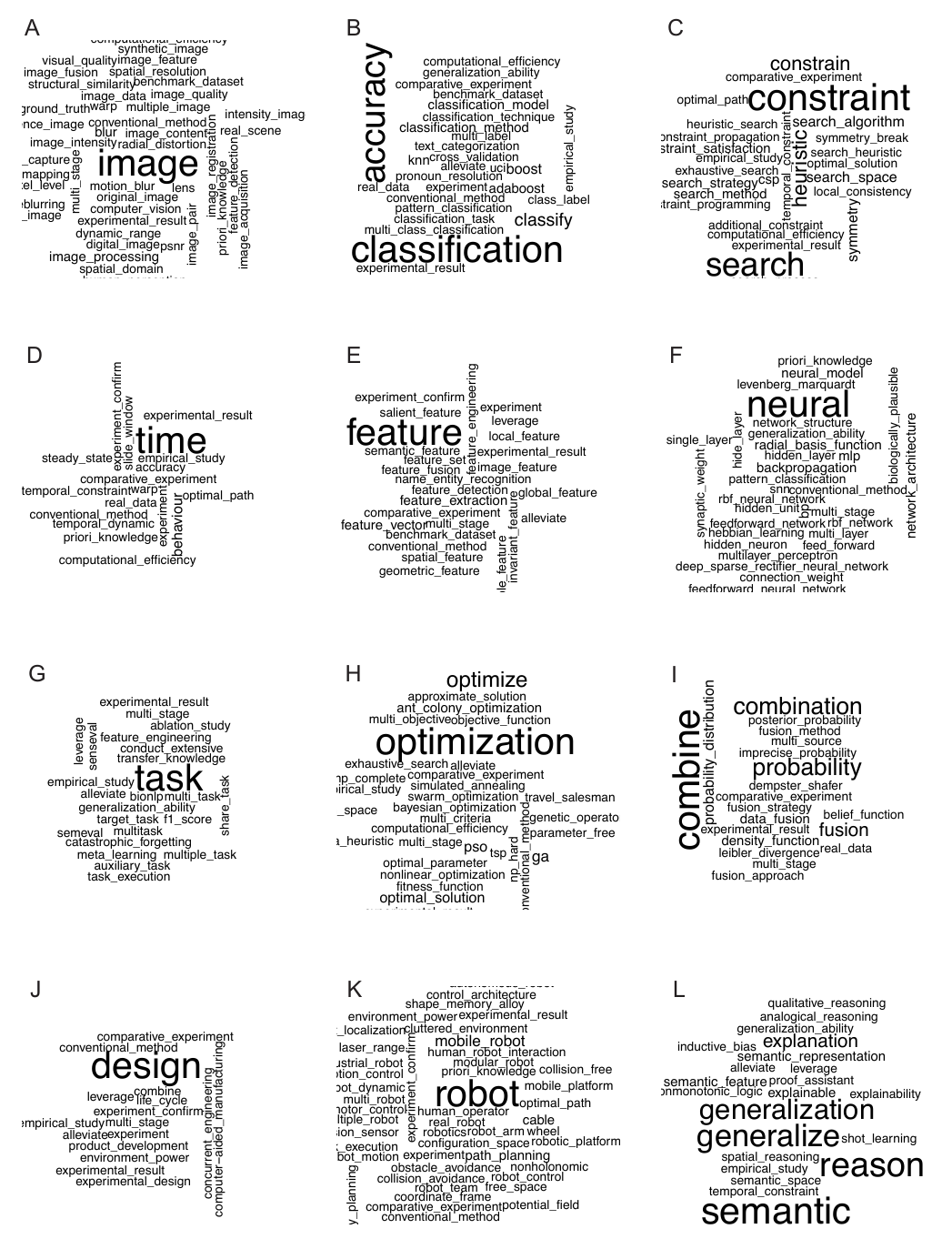}
\caption{Word cloud of the top 12 largest topics in the corpus. The size of a word is based on its FREX score within the topic.}
\label{fig:si:cloud}
\end{figure}

\begin{figure}[t!]
\centering
\includegraphics[trim=0 6 0 0, width=\linewidth]{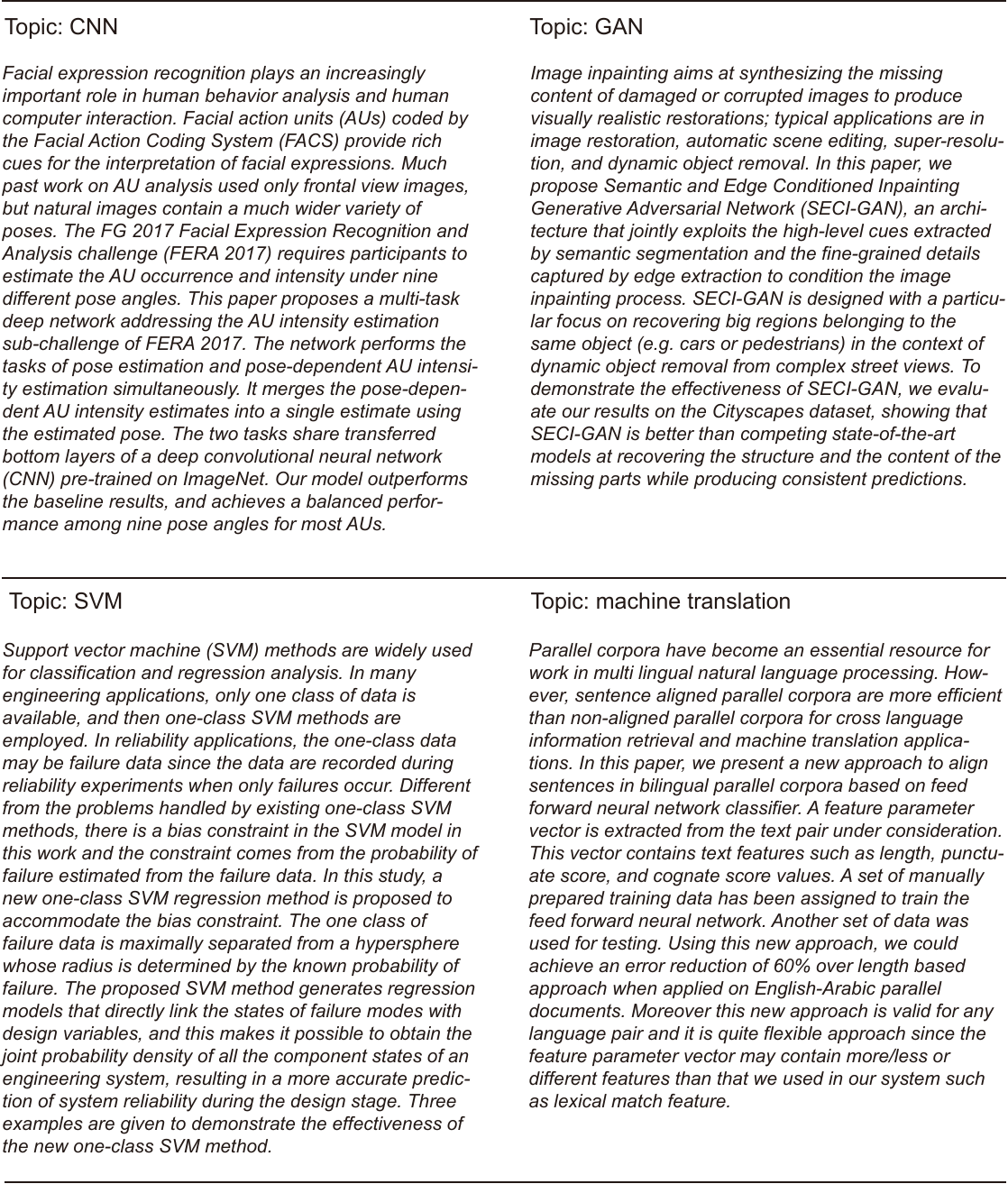}
\caption{Most representative papers of the four topics.}
\label{fig:si:case_doc}
\end{figure}

\begin{sidewaysfigure}[p]
\centering
\includegraphics[trim=0 0 0 0, width=\linewidth]{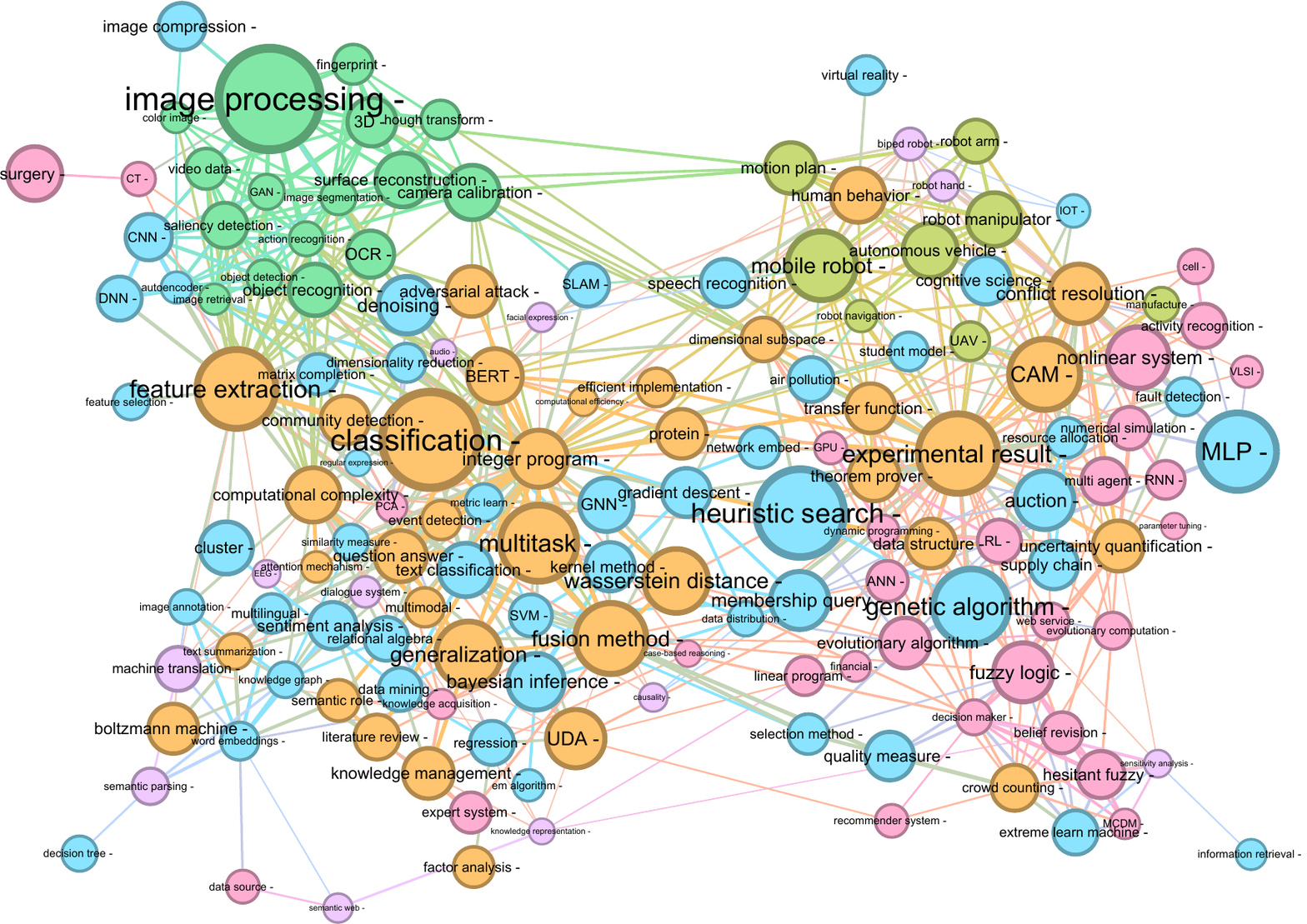}
\caption{The same as Fig~2C in the main text, with all the nodes labeled.}
\label{fig:si:label}
\end{sidewaysfigure}

\begin{figure}[t!]
\centering
\includegraphics[trim=0 6 0 0, width=\linewidth]{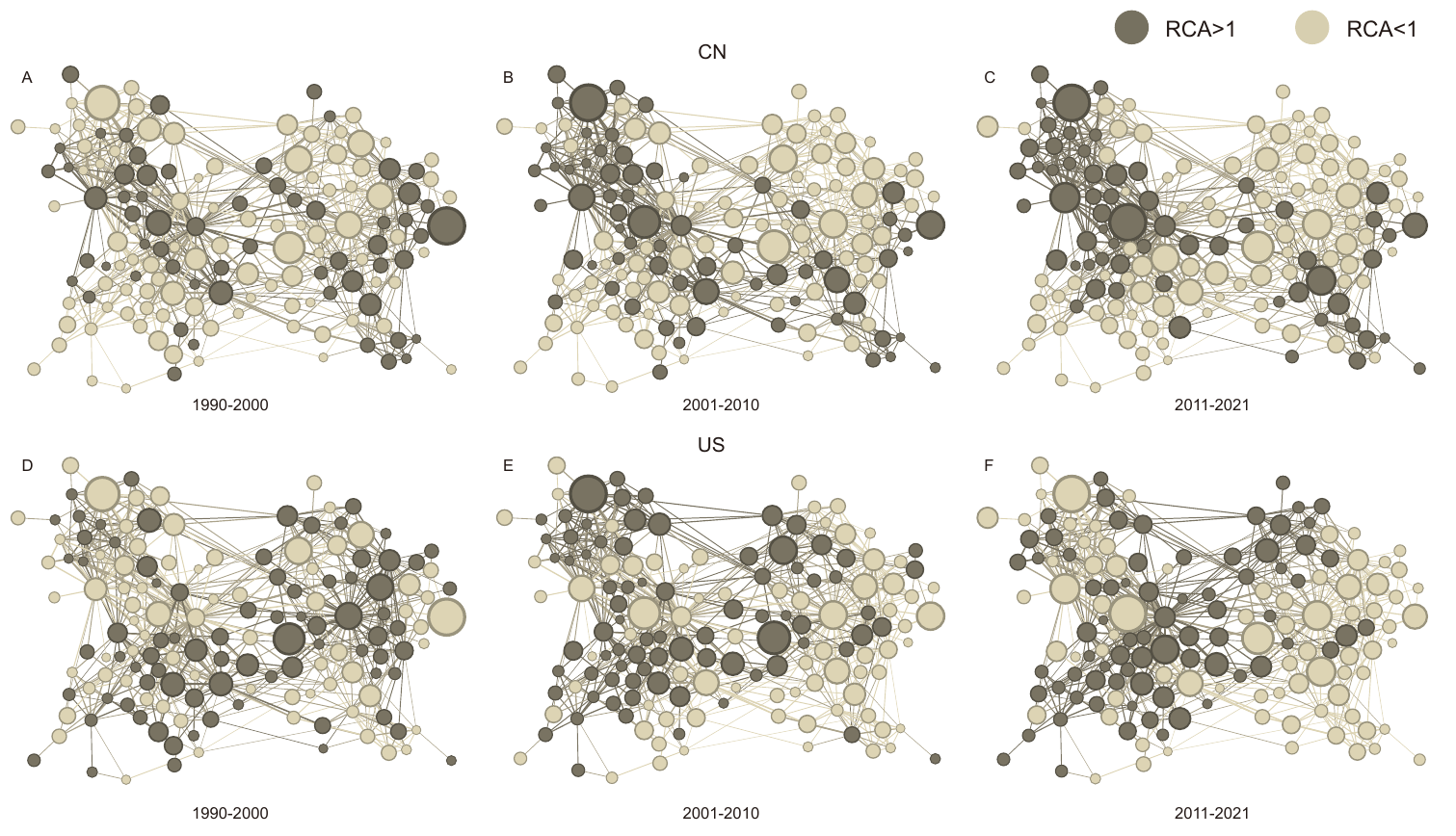}
\caption{Temporal changes in specialized topics of China and the US. Both countries exhibit a transition to a concentration in specific topics over time.}
\label{fig:si:cn_us}
\end{figure}

\begin{figure}[t!]
\centering
\includegraphics[trim=0 6 0 0, width=.9\linewidth]{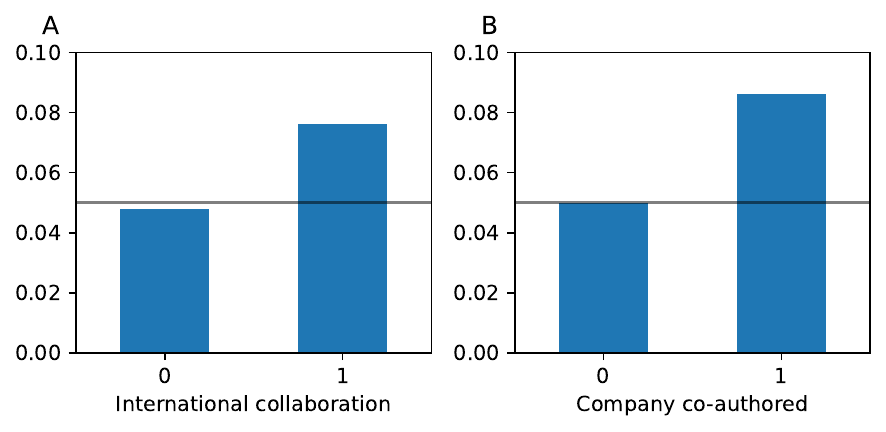}
\caption{International collaboration (A) and company coauthorship (B) are two factors associated with the production of high-impact papers.}
\label{fig:si:hit_factor}
\end{figure}

\begin{figure}[t!]
\centering
\includegraphics[trim=0 8 0 0, width=.9\linewidth]{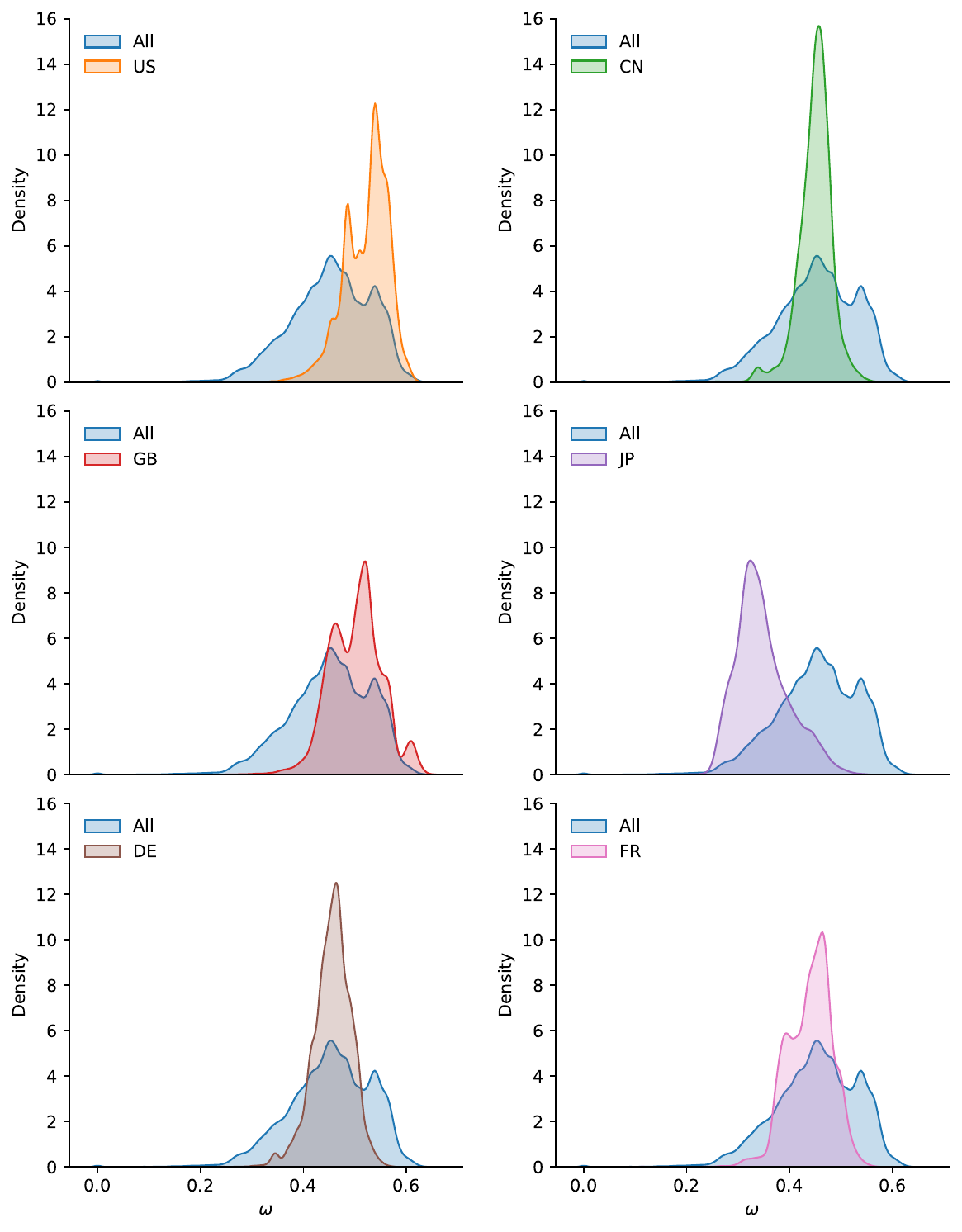}
\caption{Distributions of adjacent development level by country.}
\label{fig:si:adj_den}
\end{figure}

\begin{figure}[t!]
\centering
\includegraphics[trim=0 6 0 0, width=\linewidth]{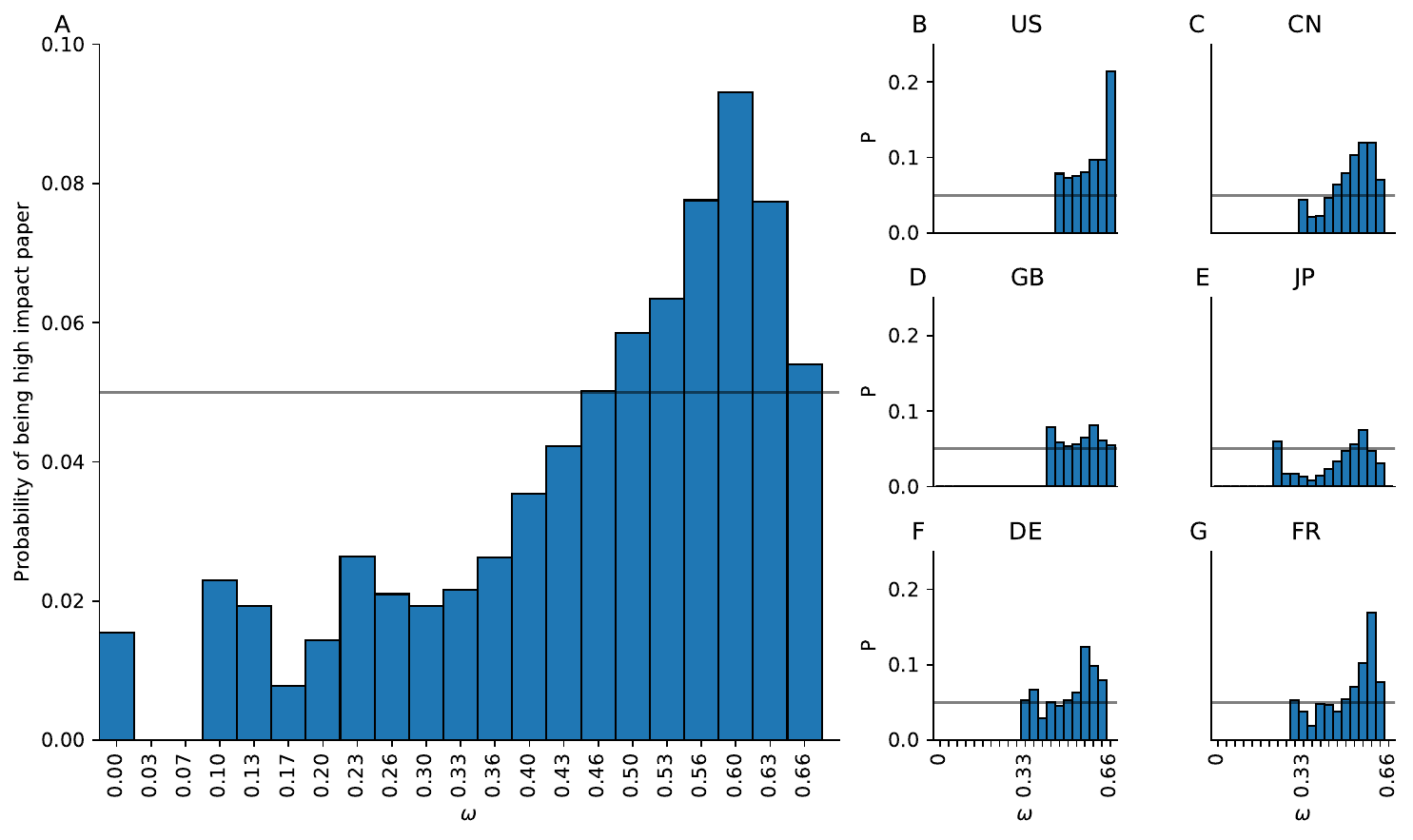}
\caption{The AI knowledge space constrains the probability of becoming high-impact papers.
(A) Countries with high adjacent development level are more likely to generate high-impact papers. 
(B--G) The impact of adjacent development level for different countries. The impact is strong for all the focused countries except the UK.}
\label{fig:si:paper-impact}
\end{figure}

\begin{figure}[t!]
\centering
\includegraphics[trim=0 8 0 0, width=\linewidth]{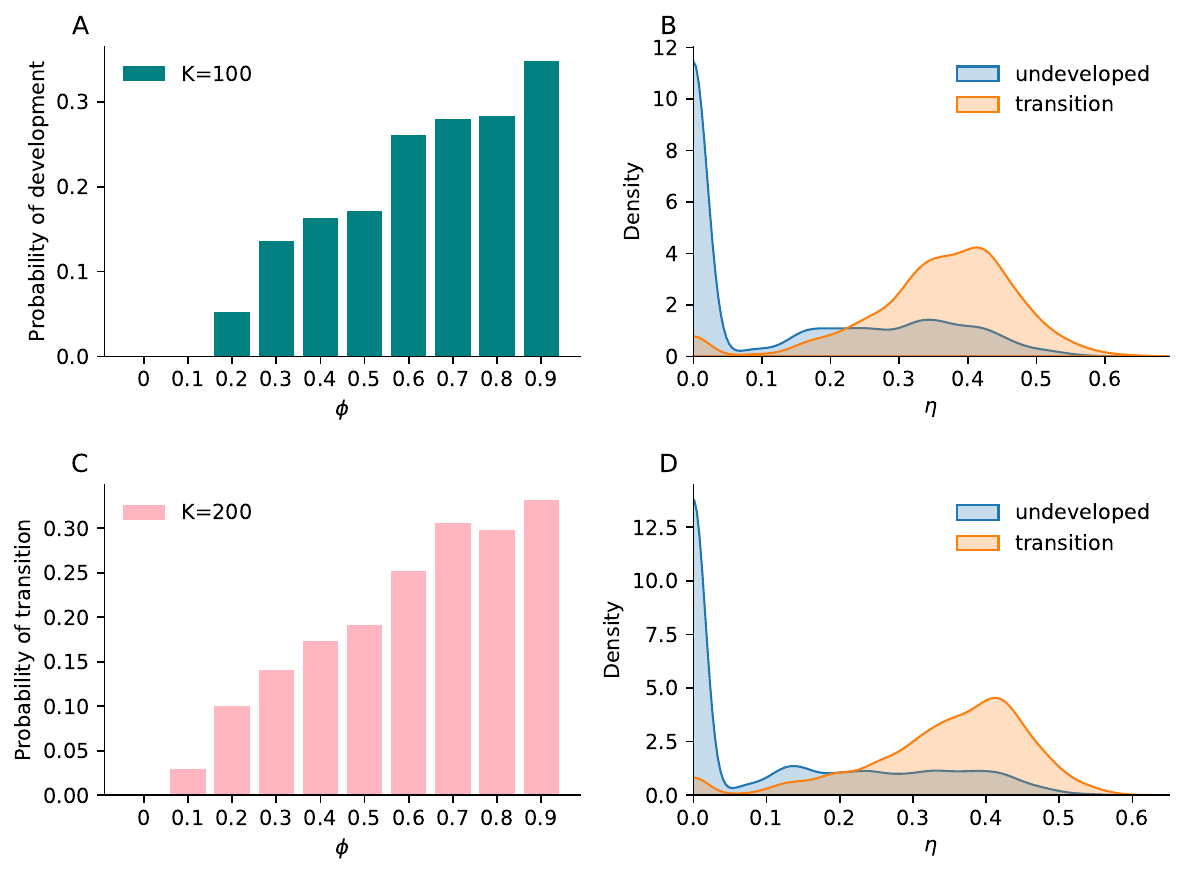}
\caption{The role of related topics in the development, when $K = 100$ (A--B) and 200 (C--D).}
\label{fig:si:trans}
\end{figure}

\begin{figure}[t!]
\centering
\includegraphics[trim=0 8 0 0, width=\linewidth]{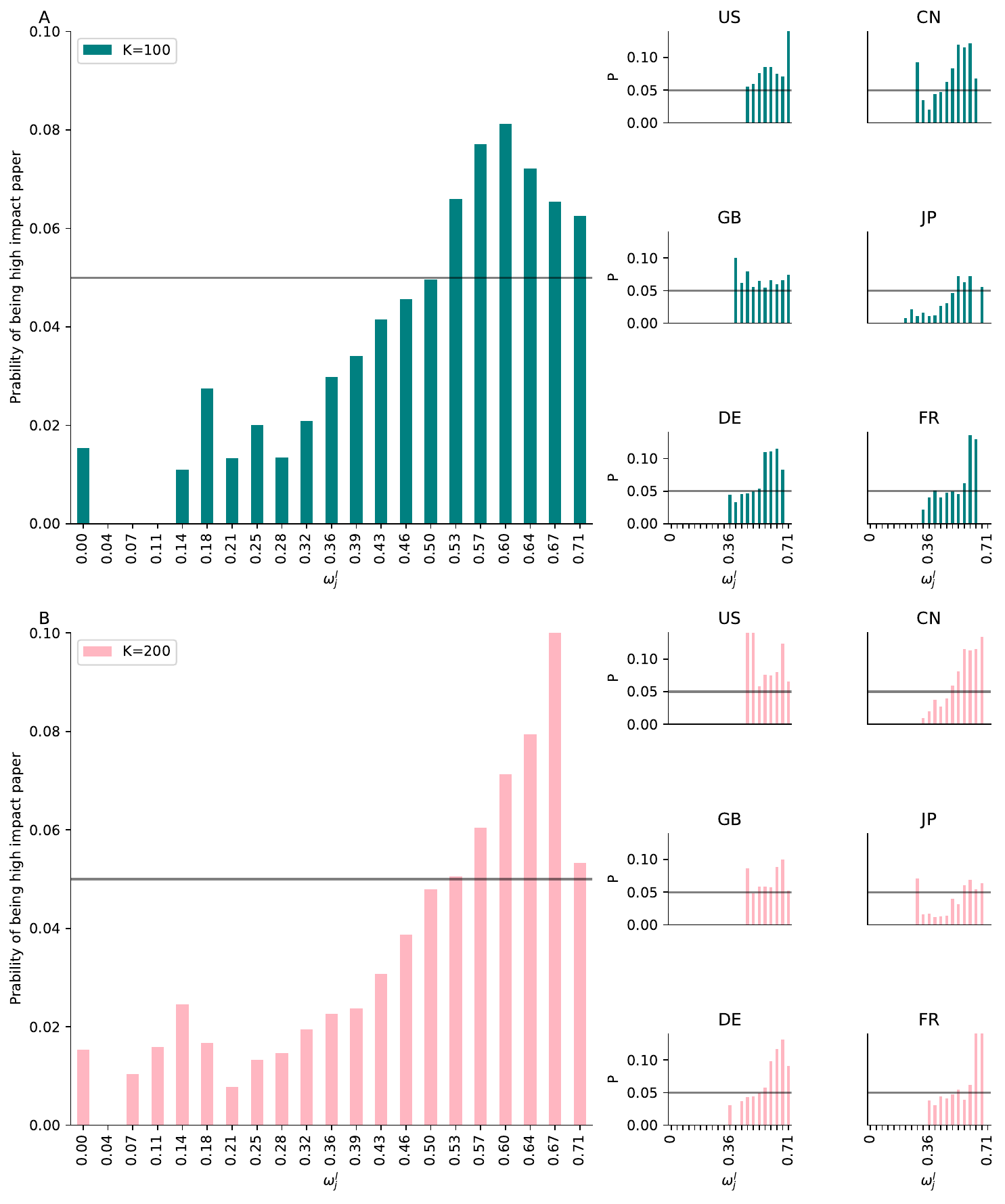}
\caption{Adjacent development level constrains the production of high-impact papers, when $K = 100$ (A) and 200 (B).}
\label{fig:si:hit_net}
\end{figure}

\linespread{1.2}
\begin{table}[t!]
\centering
\caption{Topics in each cluster.}
\label{table:topic_cluster}
\begin{tabular}{cp{13cm}}
\toprule
Cluster & Topics \\
\midrule
1 & experimental result, UDA, uncertainty quantification, factor analysis, crowd counting, dimensional subspace, conflict resolution, human behavior, feature extraction, CAM, text summarization, computational complexity, boltzmann machine, integer program, attention mechanism, multimodal, fusion method, knowledge management, BERT, transfer function, efficient implementation, theorem prover, data structure, literature review, adversarial attack, protein, community detection, semantic role, wasserstein distance, question answer, classification, multitask, generalization, event detection, computational efficiency    \\
2 & 3D, image processing, fingerprint, hough transform, GAN, video data, color image, saliency detection, surface reconstruction, image retrieval, OCR, image segmentation, object recognition, camera calibration, action recognition, object detection  \\
3 & em algorithm, word embeddings, MLP, decision tree, data distribution, information retrieval, membership query, image compression, auction, extreme learn machine, matrix completion, genetic algorithm, network embed, kernel method, IOT, feature selection, CNN, DNN, virtual reality, cognitive science, fault detection, autoencoder, data mining, image annotation, regression, cluster, speech recognition, student model, heuristic search, quality measure, metric learn, dimensionality reduction, supply chain, text classification, selection method, air pollution, similarity measure, relational algebra, denoising, bayesian inference, SLAM, SVM, knowledge graph, gradient descent, multilingual, regular expression, sentiment analysis, GNN, resource allocation  \\
4 & robot arm, autonomous vehicle, motion plan, robot manipulator, manufacture, robot navigation, UAV, mobile robot           \\
5 & semantic parsing, sensitivity analysis, semantic web, dialogue system, robot hand, audio, machine translation, causality, facial expression, knowledge representation, EEG, biped robot      \\
6 & parameter tuning, data source, RL, VLSI, MCDM, multi agent, cell, linear program, RNN, surgery, web service, evolutionary computation, hesitant fuzzy, belief revision, recommender system, dynamic programming, case-based reasoning, nonlinear system, financial, ANN, numerical simulation, expert system, CT, GPU, PCA, activity recognition, fuzzy logic, knowledge acquisition, evolutionary algorithm, decision maker   \\
\bottomrule
\end{tabular}
\end{table}

\end{document}